\documentclass[fleqn,usenatbib]{mnras}

\usepackage{newtxtext,newtxmath}

\usepackage[T1]{fontenc}

\DeclareRobustCommand{\VAN}[3]{#2}
\let\VANthebibliography\thebibliography
\def\thebibliography{\DeclareRobustCommand{\VAN}[3]{##3}\VANthebibliography}

\usepackage{graphicx}	
\usepackage{amsmath}	
\usepackage{mathtools}
\usepackage{xspace}
\usepackage{graphicx}
\usepackage{graphbox}
\usepackage{bbm}
\usepackage{soul}
\usepackage{float}
\usepackage{booktabs}
\usepackage{tabularx}
\usepackage{multirow}
\usepackage{braket}
\usepackage{threeparttable}

\newcommand{\xdesc}{\ensuremath{\mathbf{x}_\text{d}}\xspace}
\newcommand{\xroot}{\ensuremath{\mathbf{x}_\text{r}}\xspace}

\newcommand{\Xprog}{\ensuremath{\mathcal{X}_{\text{p}}}\xspace}
\newcommand{\nprog}{\ensuremath{N_\text{p}}\xspace}
\newcommand{\zdesc}{\ensuremath{z_\text{d}}\xspace}
\newcommand{\zprog}{\ensuremath{z_\text{p}}\xspace}

\newcommand{\xprogi}[1]{\ensuremath{\mathbf{x}_{\text{p}, #1}}\xspace}
\newcommand{\xsnapi}[1]{\ensuremath{\mathbf{x}^{(#1)}}\xspace}

\newcommand{\Mprogi}[1]{\ensuremath{M_{\text{p}, #1}}\xspace}
\newcommand{\Mdesc}{\ensuremath{M_{\text{d}}}\xspace}
\newcommand{\Mroot}{\ensuremath{M_{\text{r}}}\xspace}

\newcommand{\hhist}{\ensuremath{\mathcal{H}_\text{hist}}\xspace}
\newcommand{\zhist}{\ensuremath{\mathcal{Z}_\text{hist}}\xspace}

\newcommand{\florah}{\texttt{FLORAH-Tree}\xspace}
\newcommand{\satgen}{\texttt{SatGen}\xspace}
\newcommand{\gureft}{GUREFT\xspace}
\newcommand{\vsmdpl}{VSMDPL\xspace}
\newcommand{\simulation}{VSMDPL\xspace}

\newcommand{\modot}{\ensuremath{\mathrm{M_\odot}}\xspace}

\newcommand{\kpc}{\ensuremath{\mathrm{kpc}}\xspace}
\newcommand{\kpch}{\ensuremath{\mathrm{kpc}\,h^{-1}}\xspace}
\newcommand{\Mpc}{\ensuremath{\mathrm{Mpc}}\xspace}
\newcommand{\Mpch}{\ensuremath{\mathrm{Mpc}\,h^{-1}}\xspace}

\newcommand{\LCDM}{$\Lambda$CDM}

\title[Merger Tree Generative Model]{Emulating Dark Matter Halo Merger Trees with Graph Generative Models}

\author[Nguyen et al.]{Tri Nguyen,$^{1, 2}$\thanks{E-mail: trivtnguyen@northwestern.edu}
Chirag Modi,$^{3}$
Siddharth Mishra-Sharma,$^{4, 5, 6}$\thanks{Currently at Anthropic; worked performed while at MIT/IAIFI.}
{L. Y. Aaron} {Yung}$^{7}$,
\newauthor
and Rachel S. Somerville$^{8}$
\\
$^{1}$Center for Interdisciplinary Exploration and Research in Astrophysics, Northwestern University, 1800 Sherman Ave, Evanston, IL 60201\\
$^{2}$NSF-Simons AI Institute for the Sky, 172 E. Chestnut St., Chicago, IL 60611, USA\\
$^{3}$Center for Cosmology and Particle Physics, New York University, 726 Broadway, New York, NY 10003, USA\\
$^{4}$NSF AI Institute for Artificial Intelligence and Fundamental Interactions, Cambridge, MA 02139, USA\\
$^{5}$Center for Theoretical Physics, Massachusetts Institute of Technology, Cambridge, MA 02139, USA\\
$^{6}$Department of Physics, Harvard University, Cambridge, MA 02138, USA\\
$^{7}$Space Telescope Science Institute, 3700 San Martin Drive, Baltimore, MD 21218, USA\\
$^{8}$Center for Computational Astrophysics, Flatiron Institute, 162 5th Ave, New York, NY 10010, USA\\
}

\date{Accepted XXX. Received YYY; in original form ZZZ}

\pubyear{\the\year{}}

\begin{document}
\label{firstpage}
\pagerange{\pageref{firstpage}--\pageref{lastpage}}
\maketitle

\begin{abstract}
Merger trees track the hierarchical assembly of dark matter halos across cosmic time and serve as essential inputs for semi-analytic models of galaxy formation.
However, conventional methods for constructing merger trees rely on ad-hoc assumptions and are unable to incorporate environmental information. 
Nguyen et al. (2024) introduced \texttt{FLORAH}, a generative model based on recurrent neural networks and normalizing flows, for modeling main progenitor branches of merger trees.
In this work, we extend this model, now referred to as \florah, to generate complete merger trees by representing them as graph structures that capture the full branching hierarchy.
We trained \florah on merger trees extracted from the Very Small MultiDark Planck cosmological $N$-body simulation.
To validate our approach, we compared the generated merger trees with both the original simulation data and with semi-analytic trees produced using the Extended Press-Schechter (EPS) formalism.
We show that \florah accurately reproduces key merger rate statistics across a wide range of mass and redshift, outperforming the conventional EPS-based approach.
We demonstrate its utility by applying the Santa Cruz semi-analytic model (SAM) to generated trees and showing that the resulting galaxy-halo scaling relations, such as the stellar-to-halo-mass relation and supermassive black hole mass-halo mass relation, closely match those from applying the SAM to trees extracted directly from the simulation.
\florah\ provides a computationally efficient method for generating merger trees that maintain the statistical fidelity of $N$-body simulations.

\end{abstract}

\begin{keywords}
galaxies: formation -- galaxies: haloes -- (cosmology:) large-scale structure of Universe -- software: machine learning 
\end{keywords}

\section{Introduction} 

A key prediction of Lambda Cold Dark Matter ($\Lambda$-CDM) cosmology is the hierarchical formation of dark matter (DM) halos, in which massive DM halos form from the successive merger of smaller, less massive DM halos \citep[e.g.][]{1978MNRAS.183..341W, 1993MNRAS.264..201K, 2010ApJ...724..915H}.
Since galaxies form and evolve within massive DM halos, the accretion and merger histories of DM halos are intimately connected to the properties of their galaxies.
For example, galaxy mergers have been predicted by simulations to influence the growth of the central supermassive black holes and the impact of active galactic nuclei feedback \citep{2005Natur.433..604D, 2006ApJ...645..986R, 2006ApJ...641...21R, 2006ApJ...641...90R, 2006ApJS..163....1H, 2008ApJS..175..356H, 2010MNRAS.402.1693H, 2007MNRAS.380..877S, 2015MNRAS.452..575S, 2013ApJ...779..136B, 2018MNRAS.481..341S}.
Mergers can also enhance the star formation rate of galaxies via merger-driven starbursts, as demonstrated by both observations \citep[e.g.][]{2003ApJ...596L...5C, 2015MNRAS.454.1742K, 2016ApJS..222...16C} and simulations \citep[e.g.][]{2008MNRAS.384..386C, 2008ApJS..175..356H, 2016MNRAS.462.2418S}.
Additionally, they have been found to dilute the central gas-phase metallicity by triggering inflows of metal-poor gas from the outer regions, as predicted by simulations \citep[e.g.][]{2010A&A...518A..56M, 2012ApJ...746..108T, 2018MNRAS.479.3381B, 2019MNRAS.482L..55T, 2020MNRAS.494.3469B}.

The hierarchical growth of DM halos is commonly characterized by ``merger trees'', which are tree-like data structures that link progenitor halos to their descendants across cosmic times, capturing both smooth accretion and discrete mergers that shape the evolution of galaxies and large-scale structures.

Merger trees are the foundation of semi-analytic models (SAMs), which use simplified prescriptions of baryonic physics to assign physical and observable galaxy properties onto DM-only merger trees~\citep[e.g.][]{1994MNRAS.271..781C, 2000MNRAS.319..168C,1993MNRAS.264..201K,somerville1999}.
Compared to full hydrodynamical simulations, SAMs enable more extensive exploration of baryonic physics parameter space due to the significantly lower computational cost. 
As a result, insights gained from SAMs have played an important role in our understanding of modern galaxy formation theory (for a review, see \citealt{2015ARA&A..53...51S}). 

There are two main approaches for generating DM-only merger trees that represent structure formation in a chosen \LCDM\ cosmology:  (1) extracting them cosmological DM-only (or $N$-body) simulations and (2) constructing them semi-analytically using various methods based on the Extended Press-Schechter (EPS) formalism \citep{1974ApJ...187..425P, 1991ApJ...379..440B, 1991MNRAS.248..332B}.

Cosmological $N$-body simulations evolve the Universe starting from a specified set of initial conditions and track the gravitational dynamics of DM particles over cosmic time.
Halo finders \citep[e.g.][]{1985ApJ...292..371D, 2009ApJS..182..608K, doi:10.1017/pasa.2019.12, 2001MNRAS.328..726S, 2013ApJ...762..109B, 2024ApJ...970..178M} then identify bound DM structures within different snapshots, and tree construction algorithms \citep[e.g.][]{2005Natur.435..629S, 2013ApJ...763...18B, 2015MNRAS.449...49R, 2019PASA...36...28E} link halos across snapshots to form merger trees.
These simulations are generally regarded as the ``ground truth'' for structure formation, though numerical resolution and inconsistencies between different halo finders and tree reconstruction algorithms can limit their accuracy and reliability \cite[e.g.][]{2011MNRAS.415.2293K, 2013MNRAS.436..150S}.

However, simulations have limited dynamic range due to computational costs that scale steeply with simulated volume and mass resolution.
Large volumes improve sample statistics and are required for capturing rare, massive halos such as galaxy clusters ($10^{14}-10^{15} \, \modot$), while high resolution is necessary to resolve halos at galactic scales. 
Due to the hierarchical nature of \LCDM, accurate modeling of galaxy formation requires resolving halos down to masses of at least $\sim 10^{10}\, \modot$.
This challenge becomes even more pronounced at high redshifts, where halos are less massive and more abundant, requiring higher mass resolution and finer temporal sampling to accurately track their evolution. Consequently, simulations must balance resolution and volume, limiting their ability to simultaneously capture the full range of halo properties across cosmic time.

On the other hand, EPS-based methods can generate merger trees with large dynamic range and are computationally much cheaper than $N$-body simulations. 
The EPS formalism describes the conditional halo mass function, that is, the probability $p(M_2, z_2 \mid M_1, z_1)$ that a halo of mass $M_1$ at $z_1$ has a progenitor of mass $M_2$ at some earlier redshift $z_2 > z_1$.
Under the assumption that halo growth follows a Markov process, this formalism can be used (along with other assumptions) to generate merger trees via Monte Carlo techniques~\citep{1993MNRAS.262..627L, somerville_kolatt1999, zentner2007, 2008MNRAS.383..615N, 2008MNRAS.383..557P, 2015MNRAS.450.1514C, 2015MNRAS.452.1217C}.
However, standard EPS-based merger trees often show discrepancies in their mass accretion histories and branching rates when compared to simulations~\citep{2000MNRAS.316..479S, 2007MNRAS.379..689L, 2019MNRAS.485.5010B}. 
Modified algorithms, which are calibrated against simulations~\citep{2008MNRAS.383..557P, 2014MNRAS.440..193J}, provide better agreement, but even these methods generally do not 
capture the environmental correlations and long-term temporal dependencies (i.e. the connection between halo merger history and large scale environment). 

Machine learning offers promising approaches for generating fast merger trees with environmental information, yet this application remains unexplored.
Prior studies such as \citet{2022ApJ...941....7J, 2024ApJ...965..101C} focus on mapping halo properties to baryon properties given existing DM halo merger trees, rather than generating the merger trees themselves.
\citet{2022mla..confE..13T} presented a graph generative model that maps properties of a halo at $z=0$ to its progenitor graphs at $z=2$, which could in principle be adopted to generate full merger trees.
However, this approach was only demonstrated for these two redshifts, leaving unclear whether it can generate full merger trees across multiple epochs.

Only \citet{2022MNRAS.514.3692R} has previously employed an image-based generative adversarial network (GAN) to emulate merger trees.
However, GANs lack tractable likelihoods and are susceptible to mode collapse (where the generator learns to produce only a limited subset of the target distribution), thus requiring more extensive validation than those presented. 
The image-based approach also limits the ability to generate trees at arbitrary redshifts outside the training range and, more importantly, cannot enforce specific physical constraints, e.g., prohibiting arbitrary splitting of halos.

Graph-based generative models from the broader machine learning literature~\citep[e.g.][]{2019arXiv190513177L, 2023arXiv230202591L} are conceptually better suited for merger tree generation but also face their own limitations.
These methods require predetermined graph sizes, making them poorly suited for the variable branching structure of merger trees, and similarly lack mechanisms to enforce physical constraints.

Instead of emulating full merger trees all at once, a simpler approach is to first model the main progenitor branch, which traces the most massive progenitors of a halo, and then reconstruct the remaining branches in a physically consistent manner.
Following this strategy, \citet{florah} (hereafter N24) developed \texttt{FLORAH}, a generative model based on recurrent neural networks (RNNs) and normalizing flows that autoregressively models main progenitor evolution.
N24 demonstrated that \texttt{FLORAH} successfully reproduces population statistics of mass and concentration histories in cosmological N-body simulations, as well as clustering properties and assembly bias for main progenitor branches.

In this work, we extend \texttt{FLORAH} to generate complete merger trees by incorporating two key components: a classifier to predict the number of progenitors at each time step, and additional neural network modules to model the mass evolution of multiple progenitors simultaneously.
Our approach draws inspiration from autoregressive graph generative models like GraphRNN \citep{2018arXiv180208773Y}, but is specifically tailored for hierarchical tree structures like the merger histories.
We refer to the updated model as \florah.

We validate our approach by demonstrating that the generated merger trees accurately reproduce key statistical properties, such as progenitor mass distributions and merger rates.
To provide additional context for our model's performance relative to established semi-analytic methods, we further compare our results against merger trees generated using the EPS-based algorithm from \citet{2008MNRAS.383..557P} as implemented in \texttt{SatGen} \citep{2021MNRAS.502..621J}.
Lastly, we apply the Santa Cruz SAM to the generated merger trees and predict key galaxy-halo scaling relations such as supermassive black hole mass-halo mass and stellar-to-halo mass relations.

The paper is structured as follows.
Section~\ref{sec:sim} summarizes the $N$-body simulations used as training datasets.
Section~\ref{sec:ml} describes the extended \florah framework, training, and inference process. 
Section~\ref{sec:result} presents a comparison between the simulation and \florah merger trees.
Section~\ref{sec:discussion} discusses the limitations of the current model and future avenues.
Finally, Section~\ref{sec:conclusion} provides concluding remarks.

\section{Simulations} 
\label{sec:sim}

\subsection{Simulation details}

We train \florah on merger trees extracted from the Very Small MultiDark Planck box of the MultiDark simulations (\vsmdpl; \citealt{multidark}).
\vsmdpl is a DM-only ($N$-body) simulation run using a version of the \texttt{L-GADGET-2} code, which is a version of \texttt{GADGET-2} \citep{gadget, gadget2} optimized for improved performance with a larger number of particles.
The initial conditions are generated starting at redshift $z_\mathrm{init}=100$ using the Zeldovich approximation.
The simulation evolves $3840^3$ DM particles, each with a mass resolution of $9.1 \times 10^{6} \, \modot$, within a cubic volume of $(160 \, \Mpch)^3$, and uses an adaptive gravitational softening length of $\epsilon = 1-2 \, \kpch$\footnote{This corresponds to a cubic volume of $(236 \, \Mpc)^3$ and $\epsilon = 1.5-2.9 \, \kpc$ in physical units, respectively.}.
The cosmological parameters are broadly consistent with Planck 2013~\citep{planck2013}: $\Omega_m = 0.307$, $\Omega_\Lambda = 0.693$, $h = 0.678$, $\sigma_8 = 0.823$, and $n_s = 0.960$.

To demonstrate the robustness of \florah, we present additional results using alternative training simulations.
Specifically, we use the Gadget at Ultrahigh Redshift with Extra-Fine Timesteps (\gureft;~\citealt{gureft_paper}) simulations, which adopt the same cosmological parameters as \vsmdpl but are designed to probe the ultra-high-redshift Universe.
Further details on \gureft and associated results can be found in Appendix~\ref{app:gureft}.

\subsection{Merger tree construction and processing}

We extract DM halos using \texttt{Rockstar}~\citep{rockstar} and construct merger trees using \texttt{Consistent-Trees}~\citep{consistent_tree}.
Halo properties presented in this work are computed by \texttt{Rockstar} using its default configuration.
For simplicity, we do not include subhalos and remove all subhalo-related information from the merger trees. 
Once a subhalo merges into a distinct halo, it is no longer tracked, resulting in so-called isotrees, which contain only the evolution of distinct halos.

For the features of each halo, we use the virial mass, $M_\mathrm{vir}$, and the DM concentration $c_\mathrm{vir} \equiv R_\mathrm{vir}/r_\mathrm{s}$, where $r_\mathrm{s}$ is the Navarro-Frenk-White (NFW;~\citealt{1996ApJ...462..563N}) scale radius and $R_\mathrm{vir}$ is the virial radius. We adopt the virial mass definition of \citet{Bryan1998}, defined relative to the critical mass density. 
The DM concentration correlates with formation time and large-scale environment \citep[e.g.,][]{2002MNRAS.331...98V, 2006ApJ...652...71W, 2015MNRAS.452.1217C} and encodes crucial information about secondary dependencies of halo clustering and bias, such as halo assembly bias.
N24 found that including $ c_\mathrm{vir}$ is necessary for their model to accurately capture assembly bias. 
\texttt{Rockstar} computes the DM concentration by fitting the NFW density profile over the particle data.\footnote{Alternatively, the DM concentration can be estimated from the maximum circular velocity and radius using the relation in \citet{2011ApJ...740..102K}.}

We exclude low-resolution halos that \texttt{Rockstar} does not reliably identify or for which we cannot accurately measure concentrations.
At the tree level, we exclude all merger trees whose root halos contain fewer than 500 particles ($4.6 \times 10^7 \, \modot$), since halos below this limit may contain too many poorly-resolved progenitors. 
At the halo level, we remove halos with fewer than 200 particles ($1.8 \times 10^9 \, \modot$) and all of their progenitors. \citet{gureft_paper} showed that 200 particles is in general sufficient for \texttt{Rockstar} to obtain accurate estimates of $ c_\mathrm{vir}$.

Additionally, even when there are sufficient particles, \texttt{Rockstar} may fail to obtain a good fit to an NFW profile in certain cases, e.g. when a halo is undergoing a major merger or contains multiple density peaks (see also \citealt{gureft_paper}).
When this occurs, \texttt{Rockstar} caps the NFW scale radius $r_\mathrm{s}$ at $R_\mathrm{vir}$, leading to an artificial ``pile-up'' at $c_\mathrm{vir} = 1$.
To mitigate this issue, we remove all halos with $c_\mathrm{vir} < 1.1$ and their progenitors.

Merger trees from \texttt{Rockstar} and \texttt{Consistent-Trees} (and any halo finding/merger tree construction algorithm) can suffer from numerical artifacts, particularly for halos undergoing rapid accretion.
Common issues include unphysical mass loss, where halos exhibit abrupt mass decreases between snapshots due to halo tracking failures \citep[e.g.][]{2013MNRAS.436..150S}, and fragmentation, where single halos are temporarily split into multiple smaller halos before re-merging, leading to spurious detections and artificially inflated merger rates \citep[e.g.][]{2015MNRAS.450.1514C}.\footnote{Mass loss may also result from physical processes such as tidal stripping or changes in halo shapes following major mergers. However, distinguishing these from numerical artifacts can be challenging.}

A full treatment of these numerical artifacts is an active area of research \citep[e.g.,][]{2011MNRAS.415.2293K, 2013MNRAS.436..150S, 2024ApJ...970..178M} and beyond the scope of this work.
Here, we simply assess whether our model can reproduce the distributions of simulated merger trees, without additional corrections.
Some issues can be addressed in post-processing by modifying the generated merger trees rather than the training dataset.
For example, N24 corrects for mass loss in the main branch by ensuring that descendant halos are at least as massive as their progenitors.
However, issues like fragmentation typically require particle data and cannot be resolved at the merger-tree level alone.

\subsection{Training and test datasets}

The \vsmdpl simulation has a volume of $(160 \, \Mpch)^3$ and contains a total of $6 \times 10^{9}$ merger trees.
To manage this extensive data volume, we spatially divide the simulation into multiple sub-volumes for training and testing.
For training and validation, we extract approximately $300,000$ merger trees from a cubic $\sim (72\, \Mpch)^3$ sub-volume ($12\%$ of the original box) and split them 80-20 between training and validation.
As discussed in the next section, despite using this reduced sub-volume, the training process still requires $\mathcal{O}(100)$ GPU hours.
For testing, we extract approximately $50,000$ merger trees from a separate, non-overlapping $(40 \, \Mpch)^3$ sub-volume to ensure independent evaluation.

Similar to N24, we implement snapshot subsampling to reduce computational demands and help improve model generalization by preventing overfitting to specific snapshot sequences. 
For training, we randomly subsample snapshots at uniform intervals of $2-6$ snapshots, extending to $z=10$. 
When snapshots are removed, progenitor halos from retained snapshots are directly linked to descendants of their immediate descendants, maintaining causal structure.
Subsampling is performed once per tree, though multiple iterations can increase data volume when needed (e.g., Appendix~\ref{app:gureft}).
For testing, we subsample every 4 snapshots extending to $z=10$ (27 snapshots), to facilitate more straightforward comparisons. 
We discuss implications of this strategy in Section~\ref{sec:discussion}.

\section{Machine Learning Framework} 
\label{sec:ml}

\begin{figure*}[t!]
    \centering
    \includegraphics[width=0.95\linewidth]{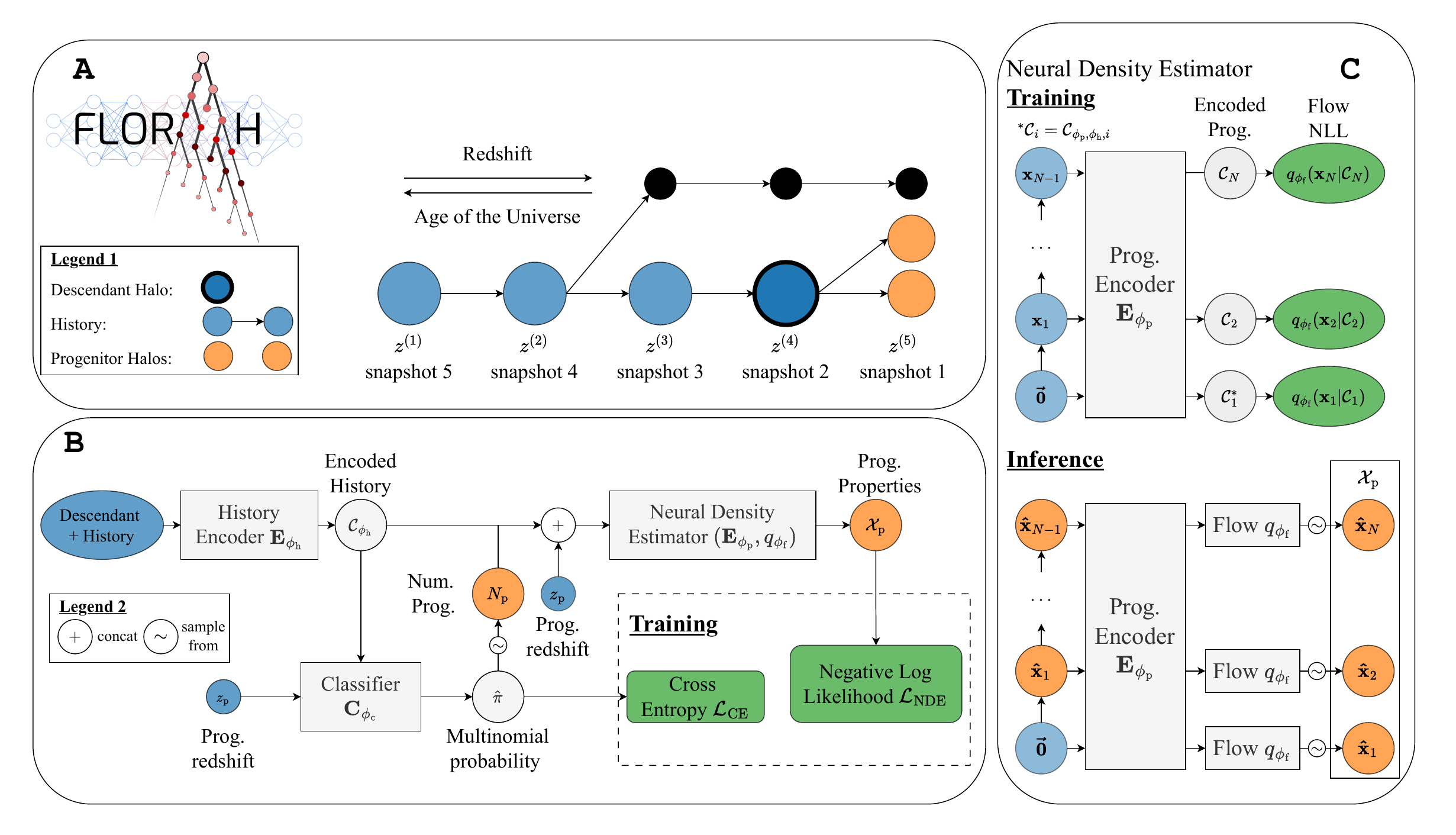}
    \caption{
    \textbf{The flowchart of \florah}.
    \textit{Panel A:} An example merger tree with a descendant halo (blue, thick border) at $z^{(4)}$ and two progenitor halos (yellow).
    Redshift increases from left to right along the branch. 
    To predict the progenitors of the descendant, \florah considers all halos in its history (blue, thin border) while ignoring halos in other branches (black).
    \textit{Panel B:} The forward model, which consists of a history encoder $\mathbf{E}_\mathrm{\phi_h}$, classifier $\mathbf{C}_\mathrm{\phi_c}$, and an NDE $\hat{q}_\mathrm{\phi_f}$.
    \florah inputs the descendant, its history, and progenitor redshift to output the number of progenitors \nprog with properties \Xprog.
    All  components are jointly optimized during training.
    \textit{Panel C:} NDE training (top) and inference (bottom).
    During training, the progenitor encoder $\mathbf{E}_\mathrm{\phi_p}$ encodes true progenitor properties as conditioning input for $\hat{q}_\mathrm{\phi_f}$, with $\mathcal{L}_\mathrm{NDE}$ summed over the sequence.
    During inference, progenitors are generated autoregressively starting with zero token $\vec{\mathbf{0}}$.}
    \label{fig:diagram}
\end{figure*}

Figure~\ref{fig:diagram} shows the schematic of \florah.
We model the assembly and merger histories of the halo \textit{backward in time}, i.e. from low to high redshift. 
By convention, ``descendant'' halos (subscript $d$) refer to halos at lower redshifts that form from the merger of ``progenitor'' halos (subscript $p$) at higher redshifts.

Given a descendant halo with a feature vector, \xdesc, at redshift \zdesc, our goal is to model the feature vectors of \nprog progenitors, $\Xprog\equiv\{\xprogi{1} \dots \xprogi{\nprog}\}$, at some redshift $\zprog > \zdesc$.
In this manner, given a root halo at the present time, we can generate its merger tree by successively predicting its progenitors at higher redshifts, until reaching a predetermined redshift or mass resolution threshold.
In this work, the feature vector \xdesc is a two-dimensional vector consisted of the halo mass and concentration, although it can include other halo properties (e.g. spin, maximum circular velocity).

To model the progenitors \Xprog, we want to also incorporate the descendant's assembly history, defined as the sequence of halos tracing back from the descendant to a single ``root'' halo at the lowest redshift. 
In the direction of increasing redshift, descendant halos split into multiple progenitor halos. 
The key observation here is that \textit{casually} and to the first approximation, these progenitor halos have already evolved independently and thus do not recombine.
As such, each descendant halo possesses a \textit{unique} assembly history that carries information about its progenitor population.
We define a history feature vector from root halo \xroot to descendant halo \xdesc as
\begin{equation}
    \hhist = \{ \xsnapi{1}, \xsnapi{2}, \dots, \xsnapi{N_\mathrm{hist}} \}, 
\end{equation}
where $\xsnapi{1}=\xroot$ and $\xsnapi{N_\mathrm{hist}} = \xdesc$, with intermediate halos ordered by increasing redshift.
We also define the redshift history vector, $\zhist=\{ z^{(1)}, z^{(2)}, \dots, z^{(N_\mathrm{hist})} \}$, which is provided as input during generation.\footnote{In practice, we use the scale factor $a = 1 / (1 + z)$ since they scale more linearly with the histories, although \zhist can be any vector that encodes the cosmological time information.}.

Since \hhist and \zhist are unique for each halo, we can write the probability distribution (PDF) of the progenitor halos as $p(\Xprog, \nprog | \hhist, \zhist, \zprog)$, where we have absorbed \xdesc and \zdesc into \hhist and \zhist, respectively. 
Following prior works (e.g. \citealt{2024arXiv240902980N, 2024arXiv240909124P}), we then decompose this distribution into product of two distributions:
\begin{align}
    \label{eq:conditional_xprop}
    p(\Xprog, \nprog  \mid \mathcal{C}) = p_\mathrm{mult}(\nprog \mid \mathcal{C}) \, p_\mathrm{prop}(\Xprog \mid \mathcal{C}, \nprog),
\end{align}
where $\mathcal{C} = (\hhist, \zhist, \zprog)$ represents the conditioning vector.
Here, $p_\mathrm{mult}$ describes the probability of having \nprog progenitors, and $p_\mathrm{prop}$ is the joint distribution of progenitor properties.

To model $p_\mathrm{mult}$, we first note that the number of progenitors is inherently discrete, which poses challenges for continuous generative models.
Assuming a maximum progenitor count, $N_\mathrm{max}$, we can model $p_\mathrm{mult}$ as a multinomial distribution, which can be optimized using a cross-entropy loss.
Alternative approaches include adding Gaussian noise to model $p_\mathrm{mult}$ as a mixture of Gaussians \citep{2024arXiv240909124P} or using normalizing flows on $\log\nprog$ \citep{2024arXiv240902980N}.
While these methods work well with large progenitor counts, they are unnecessary for our application.
Given the snapshot spacing of \simulation, we find \nprog to be typically small, with most mergers involving only 2-3 massive progenitors, making the direct multinomial approach both simple and effective.

To model $p_\mathrm{prop}$, we adopt a sequential approach similar to \citet{2024arXiv240909124P} to handle the variable number of progenitors per descendant.
Given a descendant, we order its progenitor features \Xprog in descending mass order (i.e. $\Mprogi{j} < \Mprogi{i}$ for $i < j$) and sequentially predict each new progenitor $\xprogi{i}$ using the previous progenitors $\{\xprogi{1} \dots \xprogi{i-1}\}$.
We can factor $p_\mathrm{prop}$ as follows,
\begin{equation}
    \label{eq:conditional_Xprog}
    p_\mathrm{prop}(\Xprog \mid \mathcal{C}, \nprog) = 
    \prod_{i=1}^{\nprog} p(\xprogi{i} \mid \{\xprogi{0}, \dots, \xprogi{i-1}\}, \mathcal{C}, \nprog).
\end{equation}
To initialize the sequence and predict the first progenitor, we introduce a starting token $\xprogi{0}=\vec{\mathbf{0}}$, which acts as a placeholder with no physical meaning but ensures a consistent input format.

\subsection{Neural network model}

The \florah model consists of three primary components: (1) an RNN encoder to encode the history feature and redshift vectors \hhist and \zhist, (2) a fully-connected (FC) classifier to model $p_\mathrm{mult}$, and (3) a neural density estimator (NDE) to model $p_\mathrm{prop}$.

The RNN history encoder consists of 4 GRU layers~\citep{2014arXiv1406.1078C}, each with 128 hidden units and RELU~\citep{2000Natur.405..947H} activation, that take in \hhist and \zhist and embed them into a 128-dimensional latent space.
Only the output vector from the final timestep in the sequence is used as the encoded representation.
Before passing through the GRU, \hhist and \zhist are first independently projected through FC layers, then summed together.
For simplicity, we will denote the final encoded representation as
\begin{equation}
    \label{eq:encoded_history}
    \mathcal{C}_\mathrm{\phi_h} = \mathbf{E}_\mathrm{\phi_h} (\hhist, \zhist),
\end{equation}
where $\mathbf{E}_\mathrm{\phi_h}$ is the encoder with trainable parameters $\mathrm{\phi_h}$.

The FC classifier, denoted as $\mathbf{C}_\mathrm{\phi_c}$ with trainable parameters $\mathrm{\phi_c}$, takes in this encoded representation and outputs a multinomial probability vector:
\begin{equation}
    \label{eq:classifier}
    \boldsymbol{\hat\pi} \equiv (\hat{\pi}_1, \hat{\pi}_2, \dots, \hat{\pi}_{N_\mathrm{max}}) = \mathbf{C}_\mathrm{\phi_c}(\mathcal{C}_\mathrm{\phi_h}, \zprog),
\end{equation}
where $\hat{\pi}_i$ denotes the probability of having $i$ progenitors, satisfying $\sum_{i=1}^{N_\mathrm{max}} \hat{\pi}_i = 1$ and $\hat{\pi}_i \geq 0$.
To account for the redshift of the progenitor \zprog, we project \zprog into a 128-dimensional space using a FC layer and then add the output to $\mathcal{C}_\mathrm{\phi_h}$.
The classifier consists of 4 FC layers, each with 16 hidden units and a GELU activation~\citep{2016arXiv160608415H}.

As mentioned, given a descendant, we order its progenitor features \Xprog in descending mass order and sequentially predict $\xprogi{i}$ using the previous progenitors $\{\xprogi{1} \dots \xprogi{i-1}\}$.
The NDE consists of an RNN encoder for embedding the progenitor sequence into a latent space and a conditional normalizing flow for density modeling. 

The progenitor encoder, denoted as $\mathbf{E}_\mathrm{\phi_p}$, follows the same architecture as the history encoder. 
To account for the progenitor count \nprog, we represent \nprog as a one-hot vector, project it onto a 128-dimensional latent space using a single FC layer, and then add to the encoded progenitor representation.
The final embedded representation is obtained by combining the encoded progenitor information, encoded history, and projected progenitor count, which we denote as:
\begin{equation}
    \label{eq:encoded_progenitor}
    \mathcal{C}_\mathrm{\phi_p, \phi_h, i} = \mathbf{E}_\mathrm{\phi_p}(\{\xprogi{0}, \dots, \xprogi{i-1}\}, \zprog, \nprog) + \mathcal{C}_\mathrm{\phi_h}.
\end{equation}
We use $\mathcal{C}_\mathrm{\phi_p, \phi_h, i}$ as the conditioning vector for the flow, which consists of 4 neural spline flows with monotonic rational-quadratic spline transformations~\citep{2019arXiv190604032D}, each with 8 knots and a hidden size of 128. 
We denote flow density as:
\begin{equation}
    \hat{q}_{\phi_\mathrm{f}}( \xprogi{i} \mid \mathcal{C}_\mathrm{\phi_p, \phi_h, i}),
\end{equation} 
with trainable parameters $\phi_\mathrm{f}$.

We find the inclusion of \nprog in the flow condition to be \textit{crucial} for obtaining good performance, likely due to mass conservation: \nprog determines how the descendant halo's mass is partitioned among progenitors.
Systems with many progenitors predominantly undergo minor mergers, while those with few progenitors involve major merger events.
This relationship between progenitor multiplicity and mass distribution in halos extracted from N-body simulations has been extensively studied \citep[e.g.][]{2000MNRAS.316..479S}.

Lastly, we experiment with the Transformer architecture~\citep{vaswani2023attention}.
In this manner, the history encoder functions as the Transformer encoder, while the progenitor encoder in the NDE serves as the decoder.
We also experiment with different ways to encode the redshift \zhist information such as positional sinusoidal encoding~\citep{vaswani2023attention}.
Similar to N24, we find that performance remains largely unchanged, with the GRU-based model being computationally faster.

The model used in this study consists of a total of $10^6$ trainable parameters.
The breakdown of the number of parameters of each component is summarized in Appendix~\ref{app:model}.

\begin{figure*}
    \centering
    \includegraphics[width=0.95\linewidth]{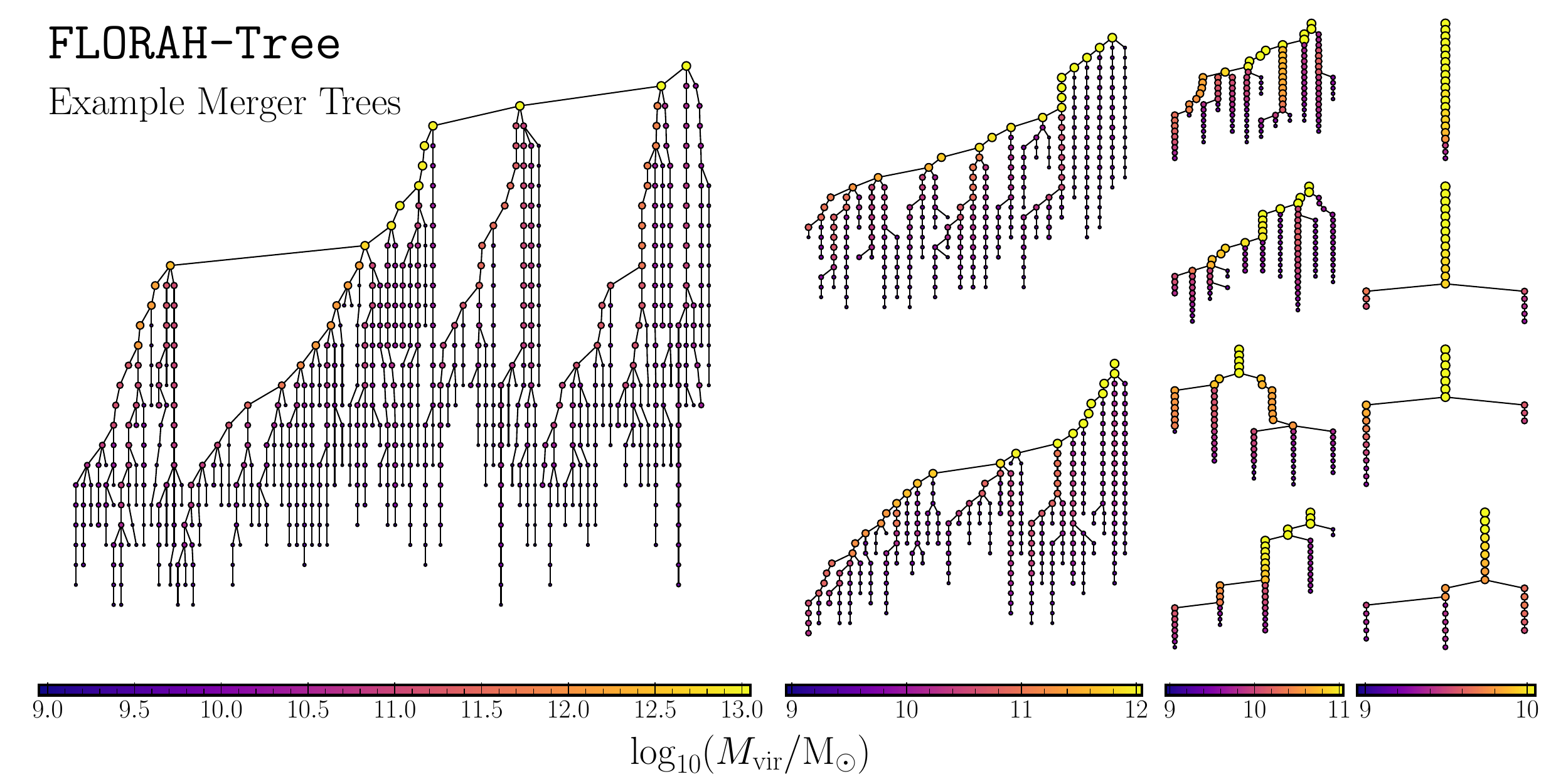}
    \caption{\textbf{Example generated merger trees.} 
    From left to right, each column shows merger trees with root masses $(10^{13}, 10^{12}, 10^{11}, 10^{10}) \, \modot$.
    Nodes in each tree represent DM halos.
    The node sizes and colors indicate the halo mass (relative within each column).
    The node vertical positions indicate redshift, increasing from top to bottom, while horizontal positions are arbitrary.
    }
    \label{fig:example_trees}
\end{figure*}

\subsection{Training}

We train all three model components simultaneously. 
Training is performed halo by halo: given a merger tree, at each training step, we select a random descendant halo and compute the corresponding loss.

Given a descendant $\xdesc$ with \nprog progenitors and true features $\{\xprogi{1} \dots \xprogi{\nprog}\}$, the classifier loss is the cross-entropy loss:
\begin{equation}
    \mathcal{L}_\mathrm{CE} = - \sum_{i=1}^{N_\mathrm{max}} \pi_{i} \log \hat{\pi}_{i} 
    = - \sum_{i=1}^{N_\mathrm{max}} \pi_{i} \log \mathbf{C}_\mathrm{\phi_c}(\mathcal{C}_\mathrm{\phi_h}, \zprog),
\end{equation}
where $\pi_i$ is the true one-hot encoding of \nprog, i.e. $\pi_i=1$ for $i=\nprog$ and 0 otherwise.
The NDE loss is the negative log-likelihood summed over all progenitors\footnote{We apply teacher forcing to ensure training stability: the model is provided with ground-truth inputs rather than its own predicted outputs during sequence generation.}:
\begin{equation}
    \mathcal{L}_\mathrm{NDE} = -\sum_{i=1}^{\nprog} \log \hat{q}_{\phi_\mathrm{f}}( \xprogi{i} \mid \mathcal{C}_\mathrm{\phi_p, \phi_h, i}).
\end{equation}
The total loss function is:
\begin{equation}
    \mathcal{L}_\mathrm{total} = \mathcal{L}_\mathrm{CE} + \alpha \mathcal{L}_\mathrm{NDE}.
\end{equation}
where $\alpha$ controls the relative weight between the losses. 
We find that $\alpha=1$ works well.

To increase training efficiency, at each training step, instead of selecting a single halo from a merger tree, we select a random branch and compute the loss for all descendant halos within that branch. 
This approach leverages the sequential nature of the GRU layers more effectively.
It also naturally places more weight on halos at lower redshifts, as halos at these redshifts appear multiple times for a single merger tree. 
This is important for autoregressive models since, during generation, any errors at early time steps will propagate to later time steps.

We use an AdamW optimizer~\citep{adamw2019, kingma2014adam} with a peak learning rate $10^{-5}$ and weight decay coefficient $0.01$.
The learning rate follows a cosine annealing schedule~\citep{2016arXiv160803983L}, with $25,000$ warm-up steps and $500,000$ decay steps.
We use a batch size of $128$.
Training converges after approximately 72 hours on a single NVIDIA Tesla A100 GPU.

\subsection{Inference}

Given a list of root halos and redshifts, generation proceeds redshift by redshift: we process all halos within each snapshot before proceeding to the next, while tracking all descendant-progenitor connections.
This enables efficient batching, with consistent input dimensionality despite the varying halo count per snapshot.
For each descendant, we encode its history, then randomly sample the progenitor count from the classifier's multinomial distribution $\hat{\mathbf{\pi}}$ and progenitor features from the NDE $\hat{q}_{\phi_\mathrm{f}}$.
After iterating over all redshifts, we reconstruct the trees using the recorded descendant-progenitor connections.

We terminate progenitor branches when their sampled mass falls below a minimum mass threshold of $1.8 \times 10^6 \, \modot$ (200 particles).
Additionally, while we explicitly enforce the descending mass ordering of progenitors during training, the flow may occasionally exhibit ``leakage'', producing progenitors that violate this condition. 
If sampled progenitors violate this condition, we simply resample, though, in practice, we find such occurrences extremely rare with sufficient training data.

For testing, we take the 50,000 root halos in the test set and randomly generate a distinct merger tree realization for each halo.
The full generation process takes about $16$ minutes on a single GPU, with reconstruction accounting for $6$ minutes.
The overall computational cost remains significantly lower than that of $N$-body simulations, and the process can be parallelized across multiple GPUs to further improve efficiency.

\section{Results} 
\label{sec:result}

We compare the properties of generated merger trees to those from the simulations.
Sections~\ref{sec:result:mass} and \ref{sec:result:merger_rate} examine the progenitor mass distributions and merger rates, respectively.
Section~\ref{sec:result:sam} shows the resulting scaling relations from applying the SC-SAM on the generated trees. 
Figure~\ref{fig:example_trees} presents a few examples of generated merger trees, grouped by root masses, showing that \florah can capture the structural diversity of merger trees.

\subsection{Progenitor mass distributions}
\label{sec:result:mass}

\begin{figure*}
    \centering
    \includegraphics[width=0.95\linewidth]{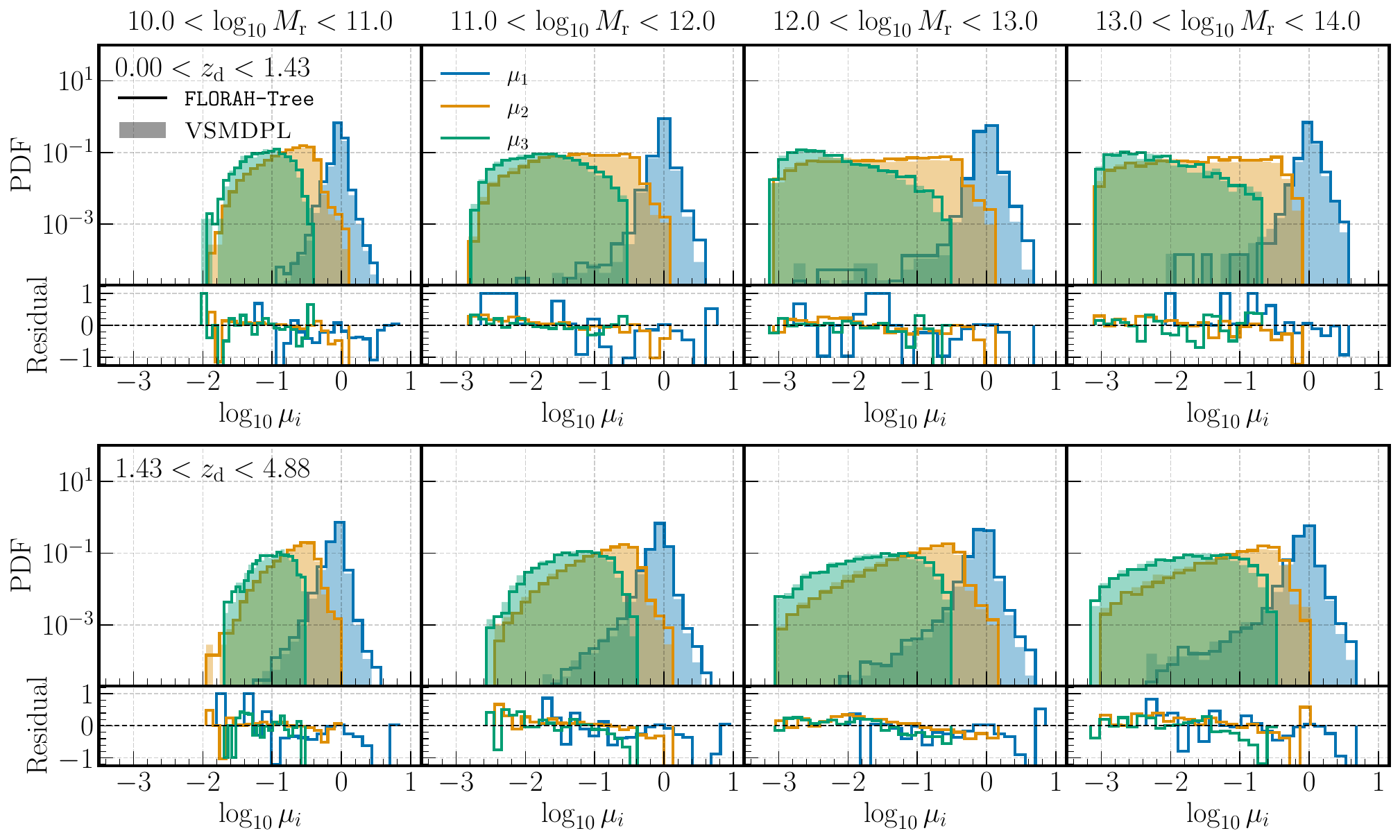}
    \caption{
    \textbf{Progenitor-descendant mass ratios.} 
    Each panel shows a different root mass \Mroot bin (row) and descendant redshift \zdesc bin (column) for ratios $\mu_i \equiv \Mprogi{i}/\Mdesc$ where $i=1, 2, 3$.
    The top row shows the distributions, while the bottom shows the residuals, defined as the fractional differences between the \simulation and \florah distributions relative to the \simulation values.
    Solid lines and shaded histograms represent \florah and \simulation merger trees, respectively.
    Colors correspond to $i=1$ (blue), $i=2$ (orange), and $i=3$ (green). 
    The close agreement between the shaded and open histograms demonstrates that the \florah method does an excellent job of reproducing these key progenitor-descendant statistics.
    }
    \label{fig:ratio_mprog_desc}
\end{figure*}
\begin{figure*}
    \centering
    \includegraphics[width=0.95\linewidth]{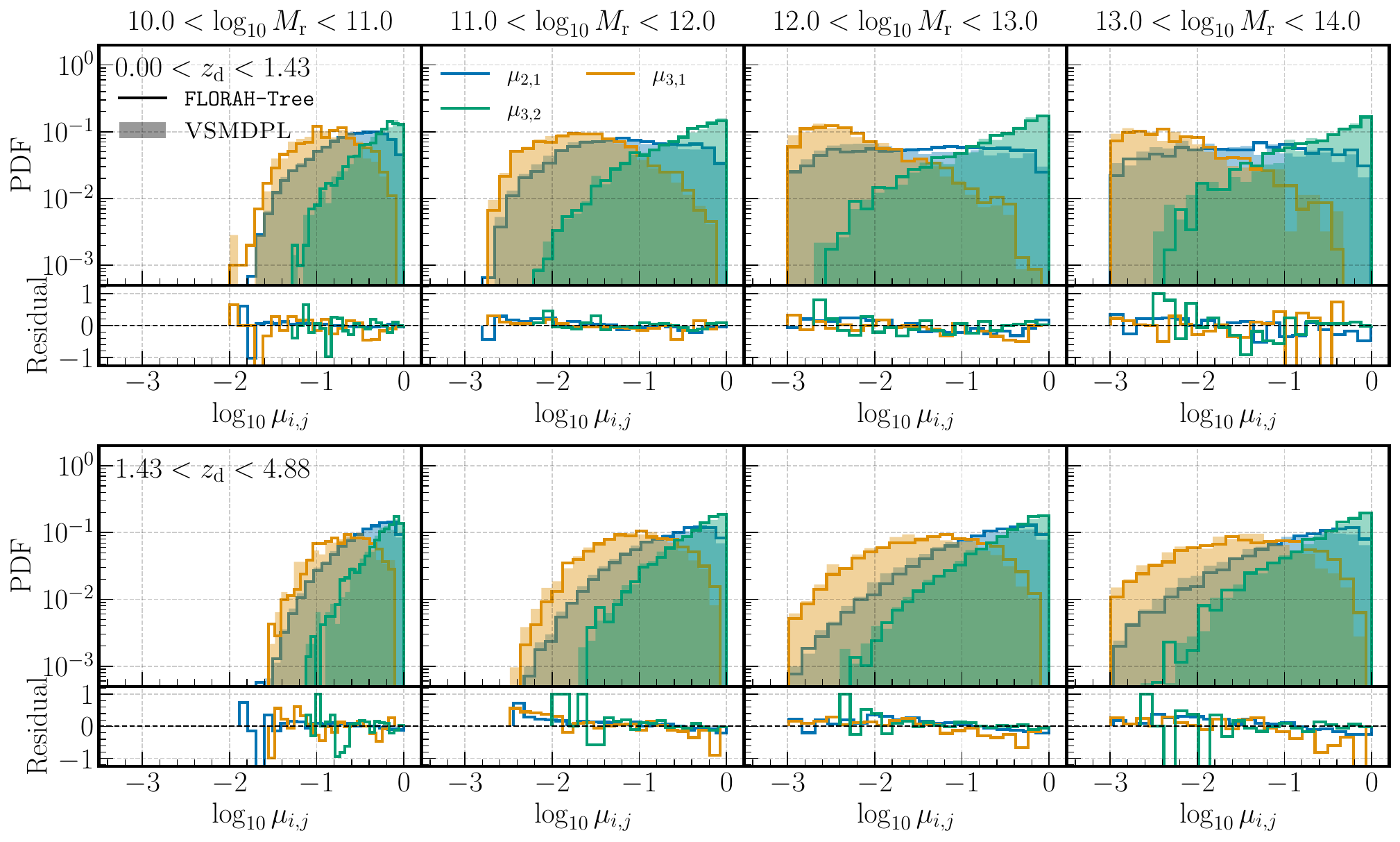}
    \caption{
    \textbf{Progenitor-progenitor mass ratios.}
   The distributions of progenitor mass ratios $\mu_{i,j} \equiv \Mprogi{i}/\Mprogi{j}$ for $i > j$.
   Panel layout matches Figure~\ref{fig:ratio_mprog_desc}.
   Colors correspond to different $(i, j)$ combinations: blue $(2, 1)$, orange $(3, 1)$, and green $(3, 2)$.
   }
    \label{fig:ratio_mprog}
\end{figure*}

We first examine the progenitor mass distributions, specifically the progenitor-descendant and the progenitor-progenitor mass ratios. 
Given a descendant halo with \nprog progenitors, we define the quantities,
\begin{equation}
    \mu_i \equiv \frac{\Mprogi{i}}{M_\mathrm{d}}; \quad  
    \mu_{i, j} \equiv \frac{\Mprogi{i}}{\Mprogi{j}}; \quad 
    i, j = 1, \dots, \nprog,    
\end{equation}
where $\Mprogi{i} > \Mprogi{j}$ for $i < j$.
Accurately reproducing the distributions of $\mu_i$ and $\mu_{i, j}$ is crucial, as inaccuracies propagate into SAMs built upon these merger trees and impact global halo properties such as the halo mass function and merger rates.

For each tree, we extract all descendant-progenitor and progenitor-progenitor pairs and put them into bins according to their root mass $\Mroot$ and descendant redshift $\zdesc$. 
We use four mass bins, with edges $(10\,, 11,\,12,\,13,\,14)\,\mathrm{\modot dex}$, and two redshift bins, with edges $(0.00,\,1.43,\,4.88)$.
Figures~\ref{fig:ratio_mprog_desc} and \ref{fig:ratio_mprog} show the distributions of $\mu_i$ and $\mu_{i, j}$, respectively for $i, j = 1, 2, 3$.

Overall, the distributions from \florah merger trees show good agreement with those from the simulations.
However, notable discrepancies emerge in the tails of the $\mu_1$ distributions, particularly for $\mu_1 > 1$, where the simulated distributions exhibit a steeper slope than our model predicts.
Interestingly, this regime represents cases where the descendant mass is smaller than that of its primary progenitor.
As discussed in Section~\ref{sec:sim}, although halo mass loss can arise from physical processes such as tidal disruption, extreme mass loss is primarily driven by numerical effects (e.g. \cite{2013MNRAS.436..150S}, which reports mass loss in as many as 30\% of halos\footnote{For our test set, halo mass loss occurs about 17\% of the time for $i=1$ (primary progenitors) and about 0.04\% for $i=2$ (secondary progenitors) across the entire dataset.}).
We speculate that these simulation-specific numerical artifacts in the $\mu_1 > 1$ regime generate inconsistent noise signatures that vary across mass scales and redshifts, complicating the model's ability to achieve precise predictions in this parameter space.

Despite this challenge, \florah reproduces the general shape of both the $\mu_i$ and $\mu_{i, j}$ distributions quite well.
Most notably, the model demonstrates consistent performance across a wide mass and redshift range.
This is particularly significant given (1) the mass imbalance in the training dataset (about 0.7\% of merger trees in the training dataset have $\Mroot > 10^{12} \, M_{\odot}$) and (2) the potential for error accumulation seen in autoregressive techniques. 

\subsection{Merger rates}
\label{sec:result:merger_rate}

\begin{figure}
    \centering
    \includegraphics[width=\linewidth]{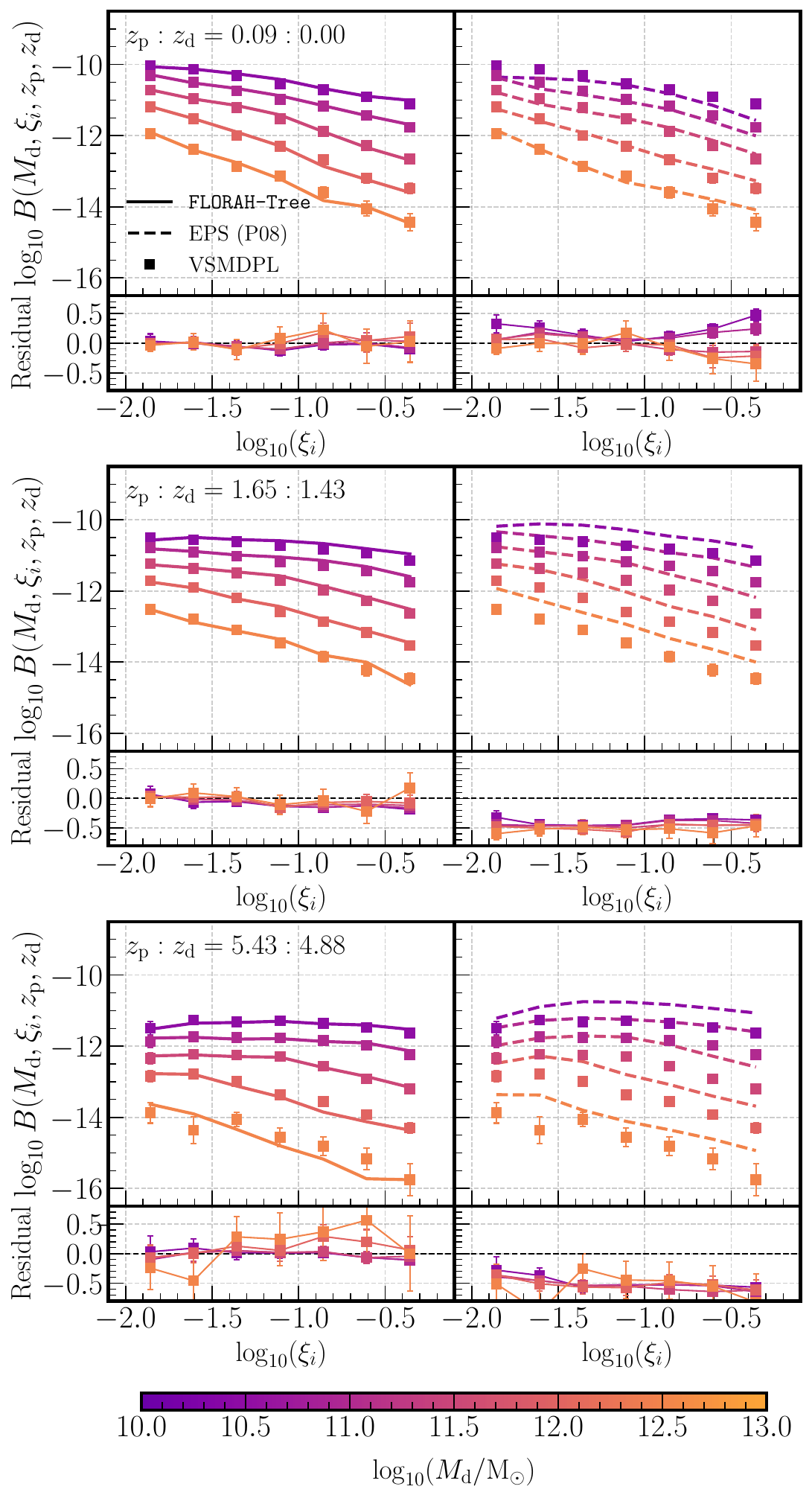}
    \caption{
    \textbf{Comparison between \florah and EPS-tree merger rates.}   
    The merger rate $B(\Mdesc, \xi_i, \zprog, \zdesc)$ is plotted as a function of the descendant mass, progenitor mass ratio, and redshifts. 
    Each row shows a different progenitor-descendant redshift slice, with each color showing a different descendant mass bin.
    In each row, the top panels show the merger rate, and the bottom panels show the residuals, $\log_{10} (B_\mathrm{sim} /B_\mathrm{gen})$.
    Error bars represent the Poisson uncertainties.
    Left panels compare the \florah and \vsmdpl merger rates, while right panels compare the EPS-tree and \vsmdpl merger rates. 
    \florah merger trees yield excellent agreement with the merger rates measured from the N-body simulations, while the EPS-based merger trees can over-estimate the merger rate by as much as $\sim 0.5$ dex at high redshift. 
    }
    \label{fig:merger_rate}
\end{figure}

The DM halo merger rate has been the subject of many studies \citep[e.g.][]{2008MNRAS.386..577F, 2010MNRAS.406.2267F, 2015MNRAS.449...49R} and is crucial for accurately modeling merger-driven processes such as SMBH growth and AGN activity \citep[e.g.][]{2005Natur.433..604D, 2006ApJ...641...21R, 2006MNRAS.365...11C, somerville2008a, 2008ApJS..175..356H}, and galaxy structural and morphological transformation and starbursts \citep{2008MNRAS.384..386C,2010MNRAS.402.1693H, 2016MNRAS.462.2418S}.

Following previous conventions from \cite{2008MNRAS.386..577F, 2010MNRAS.406.2267F}, we define the merger rate in terms of the descendant halo mass and the mass ratio of each progenitor relative to the most massive progenitor. 
For consistency with the notation in these studies, we explicitly define the mass ratio of a progenitor to the first progenitor as $\xi_i \equiv \mu_{i1}$, where $i > 1$.

We first examine the merger rates as a function of descendant mass, progenitor mass ratio, and redshift. 
Specifically, we define the ``volumetric merger rate'' $B(\Mdesc, \xi_i, \zprog, \zdesc)$ as the number of mergers per unit comoving volume, per unit descendant mass, per unit mass ratio, and per unit redshift interval, for systems where the descendant halo has mass $\Mdesc$ at redshift $\zdesc$, formed from progenitors with mass ratio $\xi_i$ at redshift $\zprog$.
For clarity, the units of $B$ are given by:
\begin{equation}
    \left[B\right] = \frac{\text{number of mergers}}{\mathrm{Mpc}^{3} \modot \Delta z \Delta \xi_i}.
\end{equation}

To provide additional context for our model's performance, we also compare \florah predictions against those from merger trees generated using the Extended Press-Schechter (EPS) formalism.
Specifically, we employ the implementation of \cite[][hereafter P08]{2008MNRAS.383..557P} as provided in the \texttt{SatGen} package \citep{2021MNRAS.502..621J}.
To ensure a consistent comparison, we use the same root mass distribution and redshift range for both the \florah and P08 EPS trees, and apply identical constraints: a minimum halo mass of $200\, M_{\mathrm{dm}}$ ($1.8 \times 10^{9} \, \modot$) and a maximum of $N_{\text{prog}} = 3$ progenitors per halo.

Figure~\ref{fig:merger_rate} compares the merger rates $B(\Mdesc, \xi_i, \zprog, \zdesc)$ of \florah, P08 EPS, and \vsmdpl trees across three progenitor-descendant redshift slices and five descendant mass \Mdesc bins, with error bars representing Poisson uncertainties.
When computing $B$, we sum over the contributions from $i=2$ (secondary) and $i=3$ (tertiary) progenitors.
To improve bin statistics, we use overlapping mass bins with lower edges $(10.0, \, 10.5, \, 11.0, \, 11.5, \,12.0)$ $\mathrm{dex}$ and a width of $2.0$ $\mathrm{dex}$ in \modot.

The results show strong agreement between the merger rates predicted by \florah and those from \vsmdpl, with consistent performance across mass and redshift bins.
Notably, the highest redshift bin primarily contains progenitors of root halos with masses exceeding $10^{12} \, M_{\odot}$, which comprise less than 1\% of the training dataset.
Accurately modeling this rare, high-mass progenitor population demonstrates the robustness of our framework.
The large scatter at the highest $z_\mathrm{desc}$ and $M_\mathrm{desc}$ bin is due to limited sample sizes.
In contrast, the EPS trees systematically overpredict merger rates.
Prior studies \citep[e.g.,][]{2008MNRAS.389.1521Z} have shown that EPS overestimates the low-redshift merger rates.
\citet{2008MNRAS.383..557P} introduced corrective factors to mitigate this issue; however, our comparison reveals that, despite showing better agreement at low redshift, the P08 EPS-based trees still overpredict merger rates by up to 50\% at high redshift. 

Lastly, we note that the $200\, M_\mathrm{dm}$ minimum mass threshold limits the completeness of minor mergers, causing systematic undercounting of mergers involving low-mass progenitors near or below this threshold.
This effect is evident in both the simulation and the generated results, particularly in the lowest \Mdesc bins and at high redshifts, where merger rates unphysically flatten and even decrease at low $\xi$.

\subsection{SC-SAM scaling relations}
\label{sec:result:sam}

\begin{figure*}
    \centering
    \includegraphics[width=0.95\linewidth]{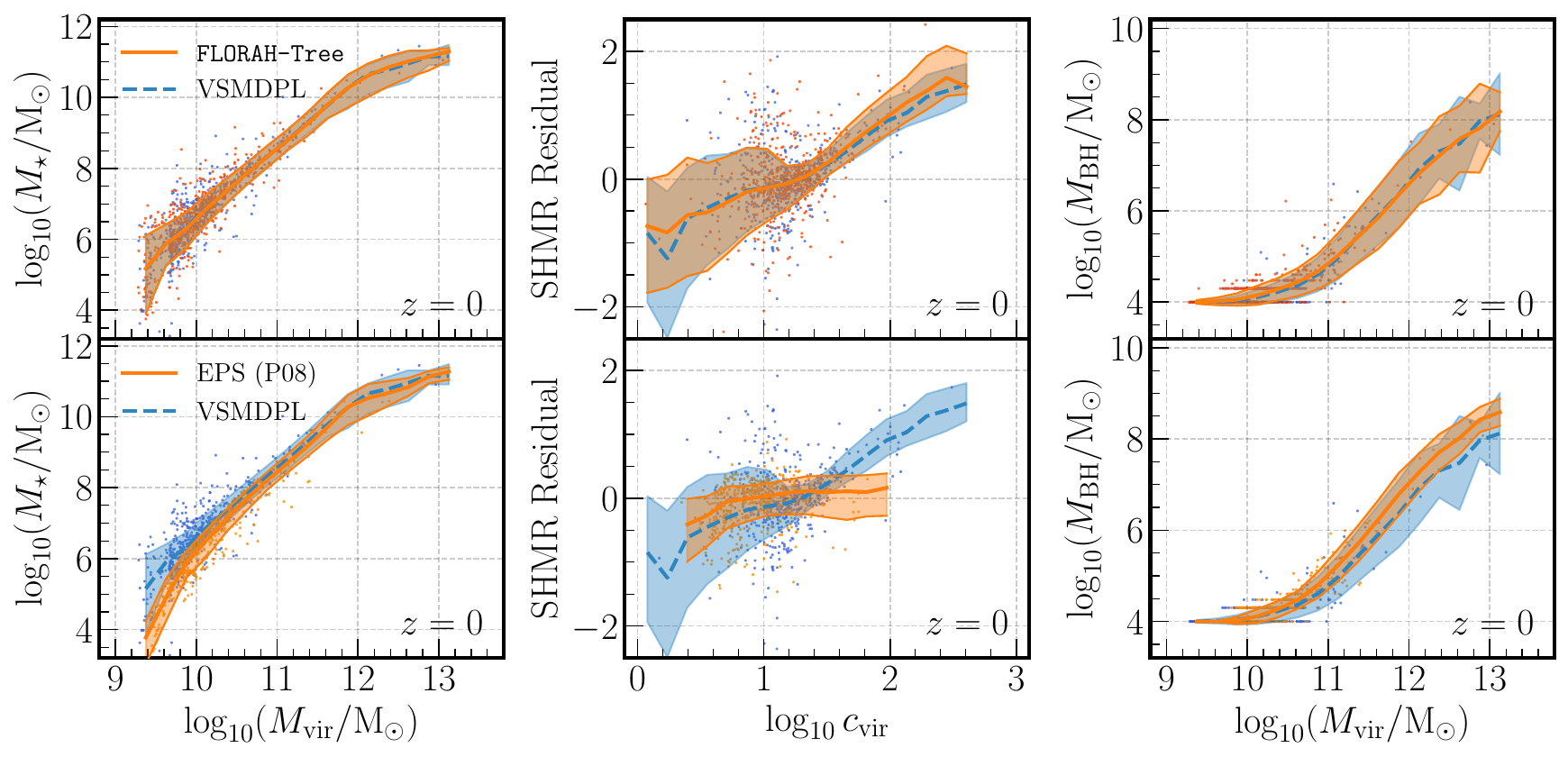}
    \caption{
        \textbf{Galaxy-halo scaling relations from the SC-SAM.}
        Comparison of galaxy-halo scaling relations predicted by the SC-SAM when implemented on top of \florah, EPS-based, and \simulation merger trees.
        The columns show (left to right): stellar-to-halo mass relation (SHMR), SHMR residual-DM concentration relation, and SMBH mass-halo mass relation at $z=0$.
        In each column, the top panel compares \florah and \simulation results, while the bottom panel compares EPS-tree and \simulation results.
        Solid lines and shaded regions indicate the medians and 68\% confidence intervals, except for the SMBH mass-halo mass relation where the means and standard deviations are shown instead.
        A subset of datapoints are displayed for clarity.
        The \florah and EPS-tree results are shown in orange, while the simulation result is shown in blue. 
        The SAM results for key galaxy scaling relations are reproduced extremely well with \florah trees, while EPS-based trees show larger discrepancies with the N-body tree based predictions. 
    }
    \label{fig:sam}
\end{figure*}

Perhaps the most stringent test of our generated trees is the predictions they yield for galaxy properties when combined with commonly used recipes for baryonic physics. 
We apply the SC-SAM~\citep{somerville2008a,10.1093/mnras/stac2297} to the generated merger trees to predict key semi-observable properties such as stellar mass and supermassive black hole (SMBH) mass.
Following Section~\ref{sec:result:merger_rate}, we compare results from applying the SC-SAM to the N-body trees, \florah, and P08 EPS merger trees.

The SC-SAM implements specific physical processes that make complete merger histories essential for accurate predictions.
The model first assigns black holes with a seed mass of $10^4 \, M_{\odot}$ to DM halos that reach a specified mass threshold.
SMBH growth occurs primarily through ``bright mode'' accretion triggered by galaxy mergers, where gravitational torques channel gas to the galactic center, simultaneously fueling rapid black hole growth and triggering starbursts.
This is supplemented by less efficient ``radio mode'' accretion from hot gas in quasi-hydrostatic halos.
For detailed descriptions, we refer the readers to \citet{somerville2008a} (see also \citealt{2012MNRAS.426..237H, 2021MNRAS.508.2706Y}).
Similarly, star formation activity occurs in two modes: episodic merger-driven starbursts and continuous quiescent star formation. 
Since both SMBH growth and a significant fraction of stellar mass assembly are directly tied to merger events, complete merger trees are essential for accurate predictions, especially in massive halos where merger histories are complex.

Since our current model predicts only the virial mass $M_\mathrm{vir}$ and DM concentration $c_\mathrm{vir}$, we limit the SC-SAM input to these two properties for all three datasets to ensure consistent comparison between \florah, P08-EPS, and \vsmdpl.
For EPS trees, \satgen computes the DM concentration of each halo from its mass assembly history, using an empirical relation that is calibrated against simulations from \citet{2009ApJ...707..354Z}.
The SC-SAM utilizes the DM concentration to calculate (1) galaxy size and (2) the tidal disruption timescale of satellite galaxies~\citep{somerville2008a}.
Although we do not include subhalos in this work, SC-SAM still evolves the satellites internally, and the galaxy properties depend on galaxy mergers, which are affected by tidal disruption.

We first examine the stellar-to-halo-mass relation (SHMR) and the SHMR residual-DM concentration relation.
The latter relation quantifies the dependency of the galaxy's stellar mass on properties beyond the halo mass and is a common probe of assembly bias \citep[e.g.][]{10.1093/mnras/stac2297}.
N24 successfully recovered both relations using the SC-SAM applied to main progenitor branches only.
Here, we investigate whether our extended \florah model, which includes the full merger tree structure with secondary and tertiary progenitors, can maintain this accuracy while capturing the additional physical processes driven by merger events.

To calculate the SHMR, we bin $M_\mathrm{vir}$ using a bin width of $0.25 \, \mathrm{\modot\, dex}$ and compute the median and 68\% confidence interval of the stellar mass $M_\star$ in each bin, excluding bins with fewer than 10 samples.
The SHMR residual is calculated as the difference between the value of $M_\star/M_\mathrm{vir}$ of each galaxy and the median value in each bin.
In Figure~\ref{fig:sam}, the left and middle columns display the SHMRs and the SHMR residual-DM concentration relations, respectively.

For both the SHMR and the SHMR residual-concentration relation, the \florah predictions align remarkably well with the \vsmdpl results, consistent with N24's findings.
The SAMs run on the P08 EPS-based trees recover the SHMR reasonably well, albeit with noticeable discrepancies at low halo masses.
However, the EPS-based trees show poor agreement with the SHMR residual-concentration relation, displaying minimal correlation between SHMR residual and concentration.
This lack of correlation highlights the well-known limitation of EPS-based merger tree algorithms in capturing the dependence of merger history on secondary halo properties, demonstrating the challenges conventional methods face when modeling galaxy formation processes that depend on halo properties beyond mass alone.

Next, we examine the SMBH mass-halo mass relation.
Since SMBH growth in the SC SAM is more sensitive to merger events than stellar mass assembly, this provides a more stringent test of the \florah\ merger trees.
As with the SHMR analysis, we bin halos by $M_\mathrm{vir}$ using a width of $0.25 \, \mathrm{dex}$ and calculate the means and standard deviations of the SMBH mass $M_\mathrm{BH}$ in each bin. 
We compute the means and standard deviations (instead of the median and 68\% interval) because the discrete seeding process creates granularity in low-mass systems with sparse merger histories.

The right column of Figure~\ref{fig:sam} compares the SMBH mass-halo mass relations for the SC SAM when run on \florah, EPS, and \vsmdpl merger trees. 
Across the full halo mass range, \florah successfully captures the relationship observed in the VSMDPL SAM run, including both the mean and scatter in the SMBH mass-halo mass relation. 
In contrast, the P08 EPS SAM run systematically overpredicts SMBH masses, consistent with its overestimation of merger rates shown in Section~\ref{sec:result:merger_rate}.
This demonstrates convincingly that our updated \florah model can effectively generate complete merger trees that capture the complex physics driving SMBH growth across cosmic time.

\section{Discussion}
\label{sec:discussion}

\subsection{Limitations of the current model}

In this work, we demonstrate that the \florah model effectively captures key merger statistics and outperforms conventional EPS-based approaches. 
The model successfully reproduces key galaxy-halo scaling relations when integrated with the SC-SAM, while incorporating environmental dependencies through halo concentration.
Here, we discuss several important limitations of the model presented in this work. 

As discussed in Section~\ref{sec:sim}, we construct merger trees by subsampling simulation snapshots at intervals of $2-6$ snapshots.
This subsampling reduces the temporal resolution for tracking galaxy formation processes.
Depending on the application, one might need higher temporal resolution to capture finer details of merger timing and halo assembly.
In future work, we aim to increase this temporal resolution further.

However, such improvements would substantially increase data volume and computational demands due to the hierarchical nature of merger trees.
The current model already requires approximately 72 hours of training on one GPU for 300,000 merger trees from a $(72\,\mathrm{Mpc}\,h^{-1})^3$ sub-volume, with a maximum of $\nprog=3$ progenitors per descendant.
These computational challenges can be addressed through advanced training methodologies.
Distributed training across multiple GPUs, mixed-precision training, and gradient accumulation techniques could significantly reduce training time and memory requirements, enabling scaling to larger volumes and finer time resolutions.
In addition, finer time steps would make events with $\nprog>2$ vanishingly rare, eliminating the need to model higher progenitor counts and reducing the required model complexity.

\subsection{Applications and Future extensions}

The current \florah framework can already enable various galaxy formation studies through its high-fidelity merger trees.
For example, lightcone construction~\citep{2021MNRAS.502.4858S, 2022MNRAS.515.5416Y, 2023MNRAS.519.1578Y} requires accurate halo assembly histories for predicting galaxy properties along the observer's past light cone, which is particularly powerful for interpreting observations from next-generation telescopes like the James Webb Space Telescope and the Nancy Grace Roman Space Telescope.
Merger trees can also be combined with models like SatGen to improve satellite galaxy population predictions within their host environments.
Beyond these applications, we identify several potential extensions that could further enhance \florah's scope and capabilities.

Cosmological simulations inherently face mass-volume trade-offs that limit their dynamic range.
A potentially powerful application involves using \florah as an ``interpolator'' between multi-tier simulations with complementary volume and mass resolutions to overcome these constraints.
Simulations such as the \gureft suite~\citep{gureft_paper} provide the complementary datasets needed for this approach.
N24 demonstrated that \texttt{FLORAH} can extrapolate mass assembly histories beyond training redshift ranges by leveraging the self-similar nature of halo growth.
Their approach combined two complementary models to generate high-resolution histories from $z=0$ to $z=25$ for root halos with masses $10^{10}-10^{14} \, \modot$, surpassing individual state-of-the-art $N$-body simulations.
Although this extrapolation capability was demonstrated only for mass assembly histories, merger rate self-similarity~\citep{2008MNRAS.386..577F, 2022ApJ...929..120D, 2025arXiv250220181J} suggests extension to complete merger trees is feasible.
This represents our highest priority for future development.

Another promising direction involves better incorporation of environmental information.
Currently, \florah captures this indirectly through halo concentration, which correlates with formation environment but lacks direct causal connections to specific initial conditions.
A natural extension would integrate \florah with initial density fields using 3D convolutional neural networks as conditional inputs, enabling generation of environmentally consistent merger trees within specific cosmic volumes.
Sufficient training data exists within the \vsmdpl volume, as our current model utilizes only a fraction of the available merger trees.

Finally, \florah could be expanded to incorporate different simulation parameters by leveraging suites with varying cosmologies and dark matter models, such as CAMELS~\citep{2021ApJ...915...71V} and DREAMS~\citep{2025ApJ...982...68R}.
This would enable applications like simulation-based inference, which requires large numbers of high-fidelity simulations across cosmological parameter space.

\section{Conclusions}
\label{sec:conclusion}

Merger trees encode the hierarchical merger and assembly history of DM halos and are a key ingredient in semi-analytic models (SAMs) of galaxy formation. 
These tree structures represent the complex evolution of DM halos, tracking both their growth through smooth accretion and discrete mergers. 

We presented \florah, an autoregressive machine learning model for generating complete hierarchical merger trees for DM halos.
This work extends the original \florah model of \citet{florah}, which only generated the main progenitor branches, to producing complete merger trees that capture the full complexity of DM halo assembly histories and discrete merger events across cosmic time.

We train \florah on merger trees extracted from a sub-volume of the \simulation box from the MultiDark suite of $N$-body simulations \citep{2011ApJ...740..102K}. 
To evaluate our method's performance, we compare the generated merger trees with those from an independent sub-volume and against merger trees generated using the Extended Press-Schechter (EPS) based algorithm proposed by \citet{2008MNRAS.383..557P} as implemented in the \texttt{SatGen} code \citep{2021MNRAS.502..621J}.
We summarize our findings as follows:
\begin{enumerate}
    \item Section~\ref{sec:result:mass} examines the progenitor-descendant (Figure~\ref{fig:ratio_mprog_desc}) and progenitor-progenitor (Figure~\ref{fig:ratio_mprog}) mass ratio distributions at several redshifts and root halo mass ranges. 
    \florah successfully models these distributions with some minor discrepancies primarily towards distribution tails.
    Notably, the model performs consistently well across the full halo mass spectrum, despite the significant imbalance in our training dataset, where root halos with masses $>10^{12} \, \modot$ represent only about 0.7\% of the total merger trees.
    Furthermore, \florah maintains accuracy at high redshifts without exhibiting the error accumulation typical of autoregressive models during sequential predictions.

    \item In Section~\ref{sec:result:merger_rate}, we compare the merger rates of \florah, EPS-based, and \vsmdpl merger trees, across a wide mass and redshift ranges (Figure~\ref{fig:merger_rate}).
    The \florah merger rates show excellent agreement with \simulation, even at the highest redshift bins, which primarily contain progenitors of the most massive root halos.
    In contrast, while the EPS-based trees reproduce merger rates moderately well at low redshifts, they overpredict merger rates at high redshifts by up to 50\%.
    
    \item In Section~\ref{sec:result:sam}, we apply the SC-SAM to the generated \florah\ merger trees and show that we reproduce the stellar-to-halo-mass relation (SHMR), SHMR residual-concentration relation, and supermassive black hole mass-halo mass relations obtained by running the SAM on the N-body trees (Figure~\ref{fig:sam}).
    EPS, on the other hand, reproduces the stellar-to-halo-mass relation fairly well, but fails to capture assembly bias effects and systematically overpredicts SMBH masses due to inflated merger rates.
    These results demonstrate that \florah  merger trees can be used as an effective basis for SAMs, offering a fast and accurate alternative to numerical cosmological simulations.

\end{enumerate}

In Section~\ref{sec:discussion}, we discuss limitations of the model presented in this work and potential future extensions for \florah.
To manage computational resources, the current model operates with reduced temporal resolution due to snapshot subsampling and is restricted to three progenitors per descendant.
Future work will extend these capabilities, and as discussed, computational challenges can be addressed through distributed training and optimized architectures.

Several promising avenues for future development include leveraging multiple simulations with complementary volumes and resolutions to expand the dynamic range beyond current $N$-body simulation capabilities, incorporating environmental information through initial density field integration, and introducing cosmological parameter conditioning using simulation suites like CAMELS and DREAMS to enable merger tree generation across diverse cosmological scenarios.

In summary, \florah successfully captures the complex hierarchical structure of merger trees across cosmic time, providing a computationally efficient alternative to $N$-body simulations for structure formation studies. 
Although this work focuses on DM halo merger trees, the \florah framework can be readily extended to generating hierarchical tree-like data structures beyond astrophysics.

\section*{Acknowledgments}
We thank Viraj Pandya and Christian Kragh Jespersen for discussions on EPS-based merger trees, and Claude-André Faucher-Giguère, Tjitske Starkenburg, Matthew Ho, Justine Zeghal, Sammy Sharief, Ronan Legin, Laurence Perreault-Levasseur for helpful feedback.

TN and SM are supported by the National Science Foundation under Cooperative Agreement PHY-2019786 (The NSF AI Institute for Artificial Intelligence and Fundamental Interactions, \url{http://iaifi.org/}).
TN is also supported by the CIERA Postdoctoral Fellowship.
RS is supported by the Flatiron Institute.
The Flatiron Institute is supported by the Simons Foundation. 
CM is supported by James Arthur Postdoctoral Fellowship at CCPP.
AY is supported by a Giacconi Fellowship from the Space Telescope Science Institute, which is operated by the Association of Universities for Research in Astronomy, Incorporated, under NASA contract HST NAS5-26555 and JWST NAS5-03127.

The CosmoSim database is a service provided by the Leibniz Institute for Astrophysics Potsdam (AIP).
The MultiDark database was developed in cooperation with the Spanish MultiDark Consolider Project CSD2009-00064.
The authors gratefully acknowledge the Gauss Centre for Supercomputing e.V. (\url{www.gauss-centre.eu}) and the Partnership for Advanced Supercomputing in Europe (PRACE, \url{www.prace-ri.eu}) for funding the MultiDark simulation project by providing computing time on the GCS Supercomputer SuperMUC at Leibniz Supercomputing Centre (LRZ, \url{www.lrz.de}).
The computations for this work were, in part, run at facilities supported by the Scientific Computing Core at the Flatiron Institute, a division of the Simons Foundation.
The data used in this work were, in part, hosted on equipment supported by the Scientific Computing Core at the Flatiron Institute, a division of the Simons Foundation.

\section*{Software}
This research makes use of the following packages:
\texttt{IPython}~\citep{PER-GRA:2007}, 
\texttt{Jupyter}~\citep{2016ppap.book...87K},
\texttt{Matplotlib}~\citep{2007CSE.....9...90H},
\texttt{NumPy}~\citep{harris2020array},
\texttt{PyTorch}~\citep{2019arXiv191201703P}, 
\texttt{PyTorch Geometric}~\citep{2019arXiv190302428F}, 
\texttt{PyTorch Lightning}~\citep{william_falcon_2020_3828935},
\texttt{SciPy}~\citep{2020NatMe..17..261V},
\texttt{SatGen}~\citep{2021MNRAS.502..621J},
\texttt{zuko}~\citep{2023zndo...7625672R},
\texttt{ytree}~\citep{ytree}.

\section*{Data Availability}

The GitHub repository for \florah can be found at \url{https://github.com/trivnguyen/florah-tree} and includes the instructions for downloading the pre-trained models and a subset of the generated merger trees.

The VSMDPL training dataset in this article uses public simulations from the CosmoSim and MultiDark database available at \url{https://www.cosmosim.org/}. 
The GUREFT training dataset will be shared on reasonable request to the
corresponding author.

\appendix

\section{Model architecture}
\label{app:model}
\begin{table}
\centering
\begin{tabular}{lc}
\hline
\textbf{Name} & \textbf{Parameters} \\
\hline
History Encoder & $4.13 \times 10^5$ \\
Neural Density Estimator & $5.70 \times 10^5$ \\
Classifier & $3.37 \times 10^4$ \\
\hline
\textbf{Total} & $\mathbf{1.02 \times 10^6}$ \\
\hline
\end{tabular}
\caption{Breakdown of the number of trainable parameters in each model component.}
\label{tab:model_params}
\end{table}

Table~\ref{tab:model_params} provides the breakdown of each model component. 

\section{Additional result on GUREFT}
\label{app:gureft}
\subsection{The \gureft simulations and dataset}

The \gureft (Gadget at Ultrahigh Redshift with Extra-Fine Timesteps;~\citealt{gureft_paper}) simulations are a suite of $N$-body simulations specifically designed to probe the assembly histories of DM halos in the ultra-high redshift universe.
The suite consists of four simulations with box sizes of $(5, \,15, \,35,\, 90) \; \Mpch$ respectively\footnote{This corresponds to $(7.37, \,22.1, \,51.6, \,133) \; \Mpc$ in physical units.}, each containing $1024^3$ particles with mass resolutions of $(6.7 \times 10^3, \, 1.8 \times 10^5, \, 2.3 \times 10^6, \, 3.9 \times 10^7) \; \modot$.
A distinctive feature of \gureft is its unprecedented temporal resolution, with the simulations storing 171 snapshots between $z\approx 6-40$, spaced at one-tenth of the halo dynamical time.
This enables detailed reconstruction of merger histories during the earliest epochs of cosmic structure formation.

The simulations were performed using \texttt{Gadget-2} \citep{gadget, gadget2} with initial conditions generated through \texttt{MUSIC} \citep{2011MNRAS.415.2101H} at $z = 200$.
Similar to \vsmdpl, the \gureft boxes adopt the Planck 2013 cosmological parameters~\citep{planck2013}: $\Omega_m = 0.307$, $\Omega_\Lambda = 0.693$, $h = 0.678$, $\sigma_8 = 0.823$, and $n_s = 0.960$.

We identified halos using \texttt{Rockstar}~\citep{2013ApJ...762..109B} with merger trees constructed via \texttt{Consistent-Trees} \citep{2013ApJ...763...18B}.
Merger trees are pre-processed using the same pipeline described in Section~\ref{sec:sim}.

We present results for the \gureft-05 box, which contains the highest number of merger trees.
We extract approximately 35,000 merger trees from the simulation and split them $80-20$ for training/validation and testing, with a further $80-20$ split between training and validation.
To augment the training dataset, we create additional samples by randomly sub-sampling each tree at intervals of $2-6$ snapshots, generating three realizations for each original tree.
This results in approximately 100,000 merger trees in the training and validation sets in total.

We train \florah using the same neural network architecture and hyperparameters described in Section~\ref{sec:ml}, with the number of trainable parameters summarized in Table~\ref{tab:model_params}.
Training takes approximately 40 hours on an NVIDIA Tesla GPU. 
For inference, we generate merger trees beginning with the root halos of the 7,000 merger trees in the test dataset, sampling halos at every four snapshots. 
To increase statistical robustness, we generate five independent realizations for each root halo, resulting in approximately 35,000 total merger trees in the generated dataset.

\subsection{Results}

\begin{figure*}
    \centering
    \includegraphics[width=0.95\linewidth]{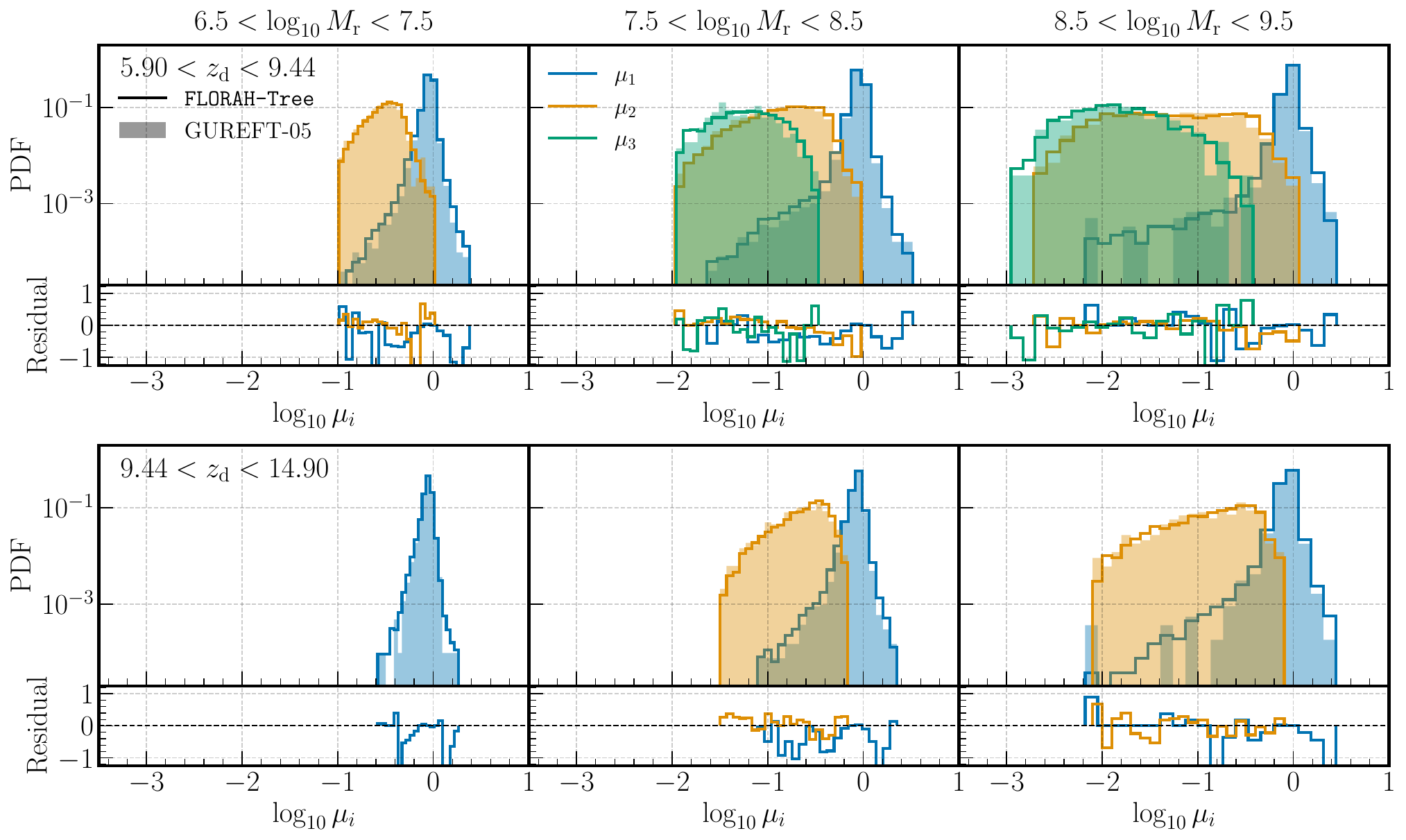}
    \caption{
    \textbf{Progenitor-descendant mass ratios.} Quantities plotted and panel layouts are the same as Figure~\ref{fig:ratio_mprog_desc}, but compare the \florah results with the \gureft-05 simulation. 
    Cases where the \gureft-05 distributions contain fewer than 100 samples are omitted.
    }
    \label{fig:app:ratio_mprog_desc}
\end{figure*}

\begin{figure*}
    \centering
    \includegraphics[width=0.95\linewidth]{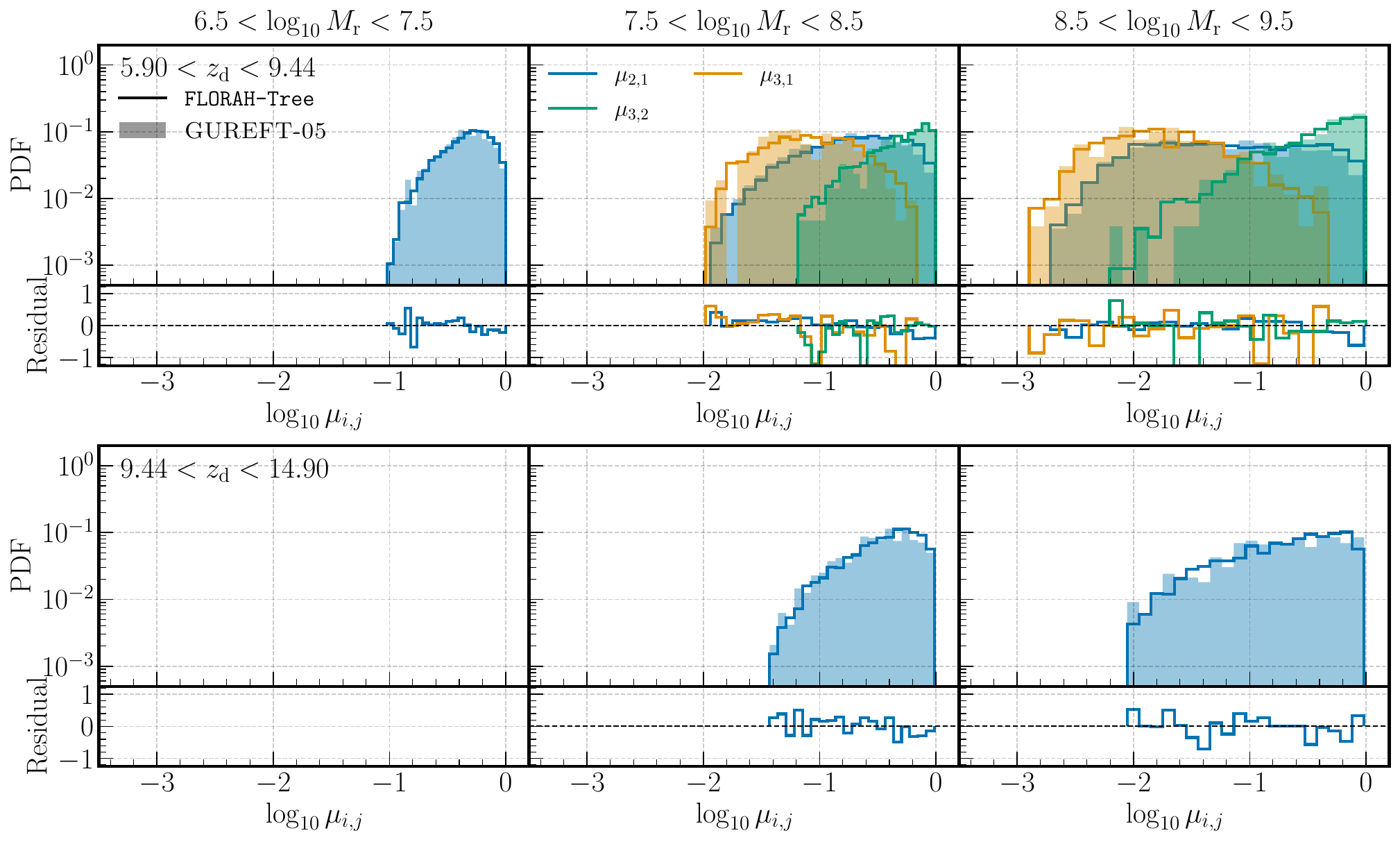}
    \caption{
    \textbf{Progenitor-progenitor mass ratios.} Quantities plotted and panel layouts are the same as Figure~\ref{fig:ratio_mprog}, but compare the \florah results with the \gureft-05 simulation. 
    Cases where the \gureft-05 distributions contain fewer than 100 samples are omitted.
    }
    \label{fig:app:ratio_mprog}
\end{figure*}

We first examine the distributions of the descendant-progenitor mass ratio $\mu_i$ and the progenitor-progenitor mass ratio $\mu_{i, j}$ for $i=1,\,2,\,3$ and $i > j$.
Figures~\ref{fig:app:ratio_mprog_desc} and \ref{fig:app:ratio_mprog} presents these distributions, respectively, and compares the \florah results with \gureft-05.
Note that certain distributions with insufficient statistics (fewer than 100 samples) are omitted from the figures. 
Qualitatively, the performance of \florah is similar to that observed with the \vsmdpl dataset, where the overall distributions are well-reproduced but with some discrepancies towards the tails of the $\mu_i$ distributions. 
We again do not observe significant differences in the performance of \florah across different mass and redshift bins. 

\begin{figure*}
    \centering
    \includegraphics[width=\linewidth]{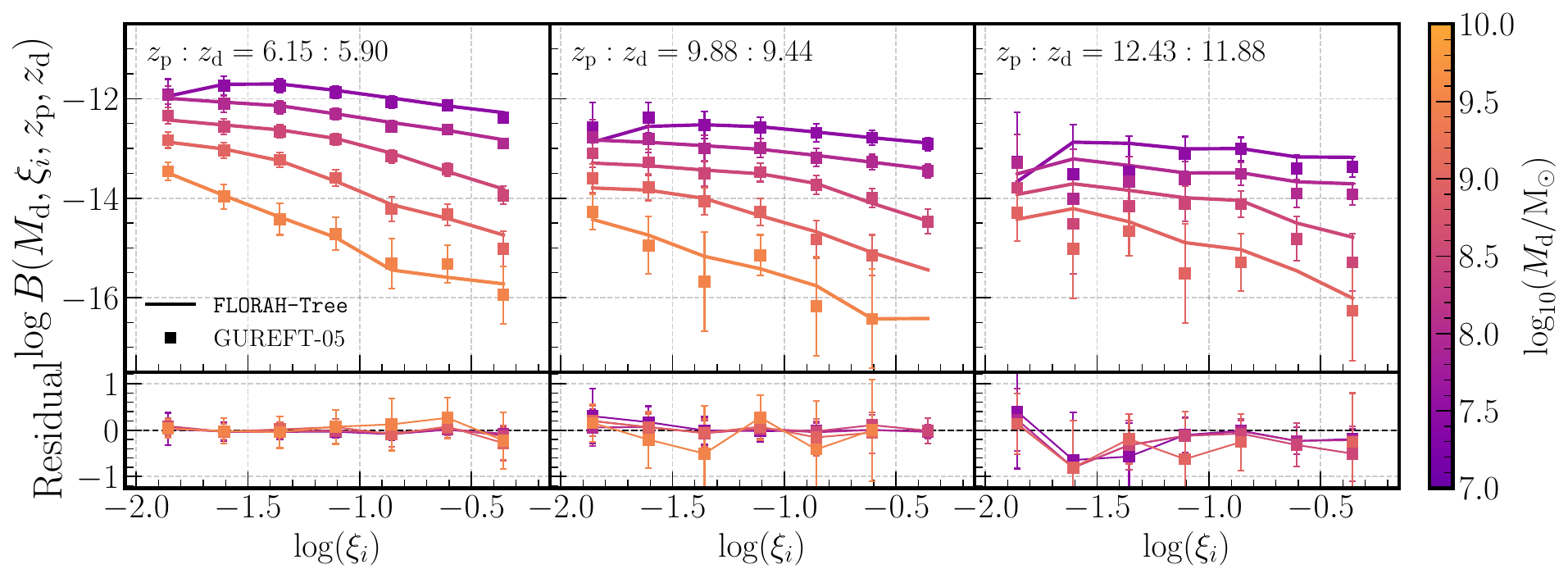}
    \caption{
    \textbf{Comparison between \florah and \gureft-05 merger rates.} 
    The merger rate $B(\Mdesc, \xi_i, \zprog, \zdesc)$ is plotted as a function of the descendant mass, progenitor mass ratio, and redshifts. 
    Each column shows a different progenitor-descendant redshift slice, with each color showing a different descendant mass bin.
    The top panels show the merger rate, and the bottom panels show the residuals, $\log_{10} (B_\mathrm{sim} /B_\mathrm{gen})$.
    The \florah and \gureft-05 results are shown as solid lines and data points, with error bars representing Poisson uncertainties.
    \label{fig:app:merger_rate}
    }
\end{figure*}

Figure~\ref{fig:app:merger_rate} shows the volumetric merger rate $B(\Mdesc, \xi_i, \zdesc, \zprog)$, which, as a reminder, is the number of mergers per unit comoving volume, per unit descendant mass, per unit mass ratio, and per unit redshift interval, for systems where the descendant halo has mass $\Mdesc$ at redshift $\zdesc$, formed from progenitors with mass ratio $\xi_i$ at redshift $\zprog$.
The error bars represent the Poisson uncertainty of $B$, calculated as the square root of the merger count (divided by the same normalization factors used for $B$).
As in the \vsmdpl case, we find that the merger rates of \florah trees are consistent with those from the \gureft-05 simulation across a wide range of descendant mass and redshift.  
At the highest redshift bin, \florah slightly overpredicts the merger rates; however, given the substantial noise in this bin, the predicted merger rates remain consistent with the simulation, with the discrepancy being within one standard deviation of the Poisson uncertainty.

\begin{figure*}
    \centering
    \includegraphics[width=\linewidth]{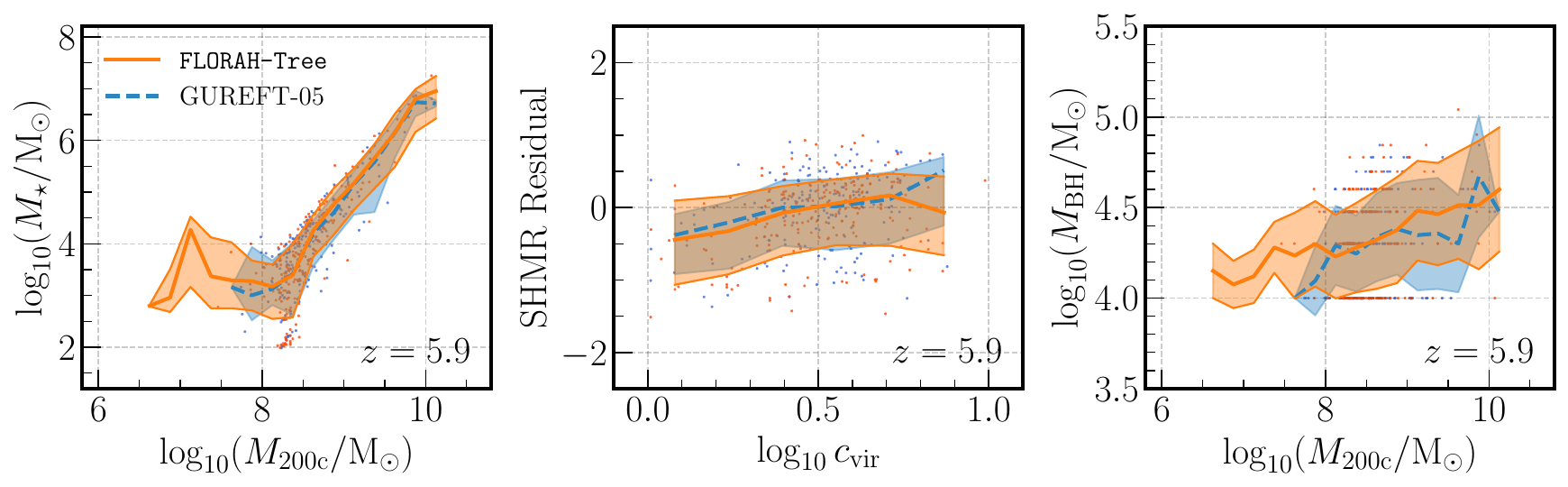}
    \caption{    
    \textbf{Galaxy-halo scaling relations by SC-SAM}. 
    From left to right, the panels show the stellar-to-halo-mass relation (SHMR), the SHMR residual-concentration relation, and the supermassive black hole (SMBH) mass-halo mass relations at $z=5.90$.
    The \florah and \gureft-05 results are shown in orange and blue respectively.
    For the SHMR and SHMR residual-concentration relation, the solid lines and bands represent the medians and 68\% confidence intervals of the relation.
    For SMBH mass-halo mass relations, they instead represent the mean and standard deviation.
    }
    \label{fig:app:sam}
\end{figure*}

Figure~\ref{fig:app:sam} presents the stellar-to-halo mass relations (SHMR), the SHMR residual-DM concentration relations (bottom panels), and  supermassive black hole (SMBH) mass-halo mass relations.
As with Section~\ref{sec:result:sam}, we observe excellent agreement between the scaling relations predicted by \florah and \gureft-05.
The scaling relations from \florah extend beyond those from \gureft-05 due to the former's larger sample size (five times more merger trees, as noted previously).

\bibliographystyle{mnras}
\bibliography{florah_bib}

\bsp	
\label{lastpage}
\end{document}